\documentclass[11pt,preprint]{aastex}
\usepackage{graphics}
\usepackage{subfigure}

\newcommand{\TN}[1]{\tablenotemark{#1}}
\newcommand{\CH}[1]{\colhead{#1}}

\shortauthors{Berg et al.}
\title{Direct Oxygen Abundances for Low Luminosity LVL Galaxies\altaffilmark{\dag}}

\author{Danielle A. Berg\altaffilmark{1}, Evan D. Skillman\altaffilmark{1}, Andrew R. Marble\altaffilmark{2,3}, Liese van Zee\altaffilmark{4}, Charles W. Engelbracht\altaffilmark{2}, 
Janice C. Lee\altaffilmark{5}, Robert C. Kennicutt, Jr.\altaffilmark{6,7}, Daniela Calzetti\altaffilmark{8}, Daniel A. Dale\altaffilmark{9}, and Benjamin D. Johnson\altaffilmark{10}}

\altaffiltext{1}{Institute for Astrophysics, University of Minnesota, 116 Church St. SE, Minneapolis, MN 55455; berg@astro.umn.edu; skillman@astro.umn.edu}
\altaffiltext{2}{Steward Observatory, University of Arizona, 933 N Cherry Ave, Tucson, AZ 85721; cengelbracht@as.arizona.edu}
\altaffiltext{3}{National Solar Observatory, 950 N Cherry Ave, Tucson, AZ 85719; amarble@nso.edu}
\altaffiltext{4}{Astronomy Department, Indiana University, 727 East 3rd Street, Bloomington, IN 47405; vanzee@astro.indiana.edu} 
\altaffiltext{5}{STScI, 3700 San Martin Drive, Baltimore, MD 21218; jlee@stsci.edu}
\altaffiltext{6}{Institute of Astronomy, University of Cambridge, Madingley Road, Cambridge CB3 0HA, UK}
\altaffiltext{7}{Steward Observatory, University of Arizona, Tucson, AZ 85721}
\altaffiltext{8}{Department of Astronomy, University of Massachusetts, 710 North Pleasant Street, Amherst, MA, 01003; calzetti@astro.umass.edu}
\altaffiltext{9}{Department of Physics and Astronomy, University of Wyoming, 1000 E. University, Laramie, WY 82071; ddale@uwyo.edu}
\altaffiltext{10}{Institut d'Astrophysique de Paris, UMR 7095, 98 bis Bvd Arago, 75014 Paris, France; johnson@iap.fr}

\altaffiltext{\dag}{Observations reported here were obtained at the MMT Observatory, a joint facility of the University of Arizona and the Smithsonian Institution.}

\begin{document}
\begin{abstract}
We present MMT spectroscopic observations of \ion{H}{2} regions in 42 
low luminosity galaxies in the \textit{Spitzer} Local Volume Legacy (LVL) survey. 
For 31 of the 42 galaxies in our sample, we were able to measure the temperature 
sensitive [\ion{O}{3}] $\lambda$4363 line at a strength of 4$\sigma$ or greater, 
and thus determine oxygen abundances using the ``direct" method. 
Our results provide the first ``direct" estimates of oxygen abundance for 19 of these galaxies.
``Direct" oxygen abundances were compared to \textit{B}-band luminosities, 4.5 $\mu$m 
luminosities, and stellar masses in order to characterize the luminosity-metallicity 
and mass-metallicity relationships at low-luminosity.

We present and analyze a ``Combined Select" sample composed of 38 objects (drawn from a sub-set of our parent 
sample and the literature) with ``direct" oxygen abundances and reliable distance determinations (based on 
the tip of the red giant branch or Cepheid variables). 
Consistent with previous studies, the $B$-band and 4.5 $\mu$m luminosity-metallicity relationships for the 38 objects were 
found to be 12 + log(O/H) = (6.27$\pm0.21) + (-0.11\pm0.01) M_{B}$ and
12 + log(O/H) = (6.10$\pm0.21) + (-0.10\pm0.01) M_{[4.5]}$ with dispersions of $\sigma$ = 0.15 and 0.14 respectively.
The slopes of the optical and near-IR L-Z relationships have been reported to be different for galaxies with luminosities greater than that of the LMC.
However, the similarity of the slopes of the optical and near-IR L-Z relationships for our sample
probably reflects little influence by dust extinction in the low luminosity galaxies.
For this sample, we derive a mass-metallicity relationship of 12 + log(O/H) = (5.61$\pm0.24) + (0.29\pm0.03) \log(\mbox{M}_{\star})$,
which agrees with previous studies; however, the dispersion ($\sigma$ = 0.15) is not significantly lower than that of
the L-Z relationships.  Because of the low dispersions in these relationships, if an accurate distance is available,
the luminosity of a low luminosity galaxy is often a better indicator of metallicity than that derived using
certain ``strong-line'' methods, so significant departures from the L-Z relationships may indicate that
caution is prudent in such cases. 
With these new ``direct" metallicities we also revisit the 70/160 $\mu$m color metallicity relationship.

Additionally, we examine N/O abundance trends with respect to oxygen abundance and B-V color.
We find a positive correlation between N/O ratio and B-V color for 0.05 $\lesssim B-V \lesssim$ 0.75:
$\log(\mbox{N/O})$ = (1.18$\pm$0.9)$\times$(B-V) + ($-$1.92$\pm$0.08), with a dispersion of $\sigma$ = 0.14, that is in agreement with previous studies.
\end{abstract}


%----------------------------------------------------------------------------------------------------------------------
%----------------------------------------------------------------------------------------------------------------------

\section{INTRODUCTION}\label{sec:intro}

There is a fundamental relationship between the mass of stars in a galaxy 
and its metallicity evolution \citep[e.g.,][hereafter, the M-Z relation]{tremonti04}. 
Empirically, this has been observed as a luminosity-metallicity relationship 
(hereafter, the L-Z relation) for low redshift dwarf galaxies
\citep[e.g.,][and references therein]{lequeux79, skillman89, lee06a} and spiral galaxies 
\citep[e.g.,][and references therein]{mccall85, garnett87, zaritsky94, tremonti04}. 
This relationship is observed over a range of 10 magnitudes in galaxy optical luminosity
\citep[e.g.,][]{zaritsky94, tremonti04, lee06a}, but the data are relatively sparse at the low 
luminosity end where the intrinsic faintness of these galaxies makes metallicity 
determinations more difficult.

The physical driver of the	M-Z relation remains under debate. 
One possibility is that low-mass galaxies are younger, in that they only recently started forming stars 
\citep{noeske2000,leitner2011}.
Another is that they have been less efficient at producing metals \citep{brooks07}.
Many studies favor a different interpretation, where supernova driven winds preferentially expel 
metals from low-mass galaxies, resulting in a lower effective yield with decreasing mass \citep[e.g.,][]{dekel86}.
However, \citet{dalcanton07} emphasizes the importance of star formation efficiency as outflows
 are an insufficient regulator in the absence of depressed star formation.
In addition, Dalcanton's calculations show that low effective yields cannot be due to gas infall.
Alternatively, \citet{koppen07} showed that the M-Z relationship may be observed naturally if a SFR-dependent, 
and therefore mass-dependent, stellar initial mass function (IMF) is assumed.
Clearly a better understanding of the mass-metallicity relationship at low-luminosity 
remains important to determine how galaxies evolve \citep[e.g., see discussion in][and references therein]{moustakas12}.
In addition, a well defined low-luminosity M-Z relationship will provide clues to the source of its measurable scatter.
While observational errors play a role, one or more physical processes may be responsible for the remainder.
 Suggestions for the scatter include variations in the star formation history \citep[e.g., recent starbursts,][]{contini02}, 
 variations in stellar surface mass density \citep{ellison08}, inflow of metal poor gas, perhaps triggered by interactions \citep{jlee2004},
 and variations in local galaxy density \citep[e.g.,][and references therein]{cooper08}.
As astronomers examine the interrelationship between chemical abundance measurements, star formation, 
gas accretion, and gas outflow by measuring the evolution of the M-Z relationship, 
a secure M-Z relationship for the current epoch is needed for comparison.

Empirical and theoretical oxygen abundance calibrations often introduce bias,
further limiting the M-Z relationship \citep[e.g.,][]{yin07, perez-montero09, moustakas10, berg11}.
Notably, for 53,000 SDSS galaxies, which span 10 orders in B-band magnitude, \citet{tremonti04} 
found a dispersion of 0.16 for their L-Z relationship and 0.10 for their M-Z relationship. 
\citet[][hereafter L06]{lee06a} were able to extend the mass-metallicity relation lower by 2.5 
decades in stellar mass using 4.5 $\mu$m luminosities for 27 nearby dwarf irregular galaxies. 
Interestingly, L06 found the dispersion in the near-infrared L-Z relationship to be smaller than 
the corresponding dispersion in the B-band L-Z relationship and 
nearly identical to that of the M-Z relationship.
The smaller dispersion in the near-infrared is not totally unexpected, 
as NIR luminosities are less sensitive to extinction from dust and variations in star formation rate.
However, the significant but uncertain stochastic effects of asymptotic giant branch (AGB) stars on the total NIR 
luminosities of low luminosity galaxies must also be considered \citep[see, e.g.,][]{fouesneau10,meidt12,
melbourne12}. 

To thoroughly examine the L-Z and M-Z relations, we need a robust sample of galaxies.
The \textit{Spitzer} Local Volume Legacy 
survey\footnote{http://www.ast.cam.ac.uk/research/lvls} \citep[LVL;][]{dale09}
covers a volume-complete sample of 258 galaxies in the local universe with 
multiwavelength observations spanning the ultraviolet to the radio. 
The LVL is leveraged by ancillary data including H$\alpha$ \citep{kennicutt08} 
and UV \citep{jlee11} imaging from the 11 Mpc H$\alpha$ and Ultraviolet Galaxy 
Survey \citep[11HUGS;][]{jlee11} and the Nearby Galaxy Survey \citep[NGS;][]{gildepaz07}. 
A subsample of the LVL also contains stellar population mapping from the ACS 
Nearby Galaxy Survey Treasury \citep[ANGST;][]{dalcanton09}, HI mapping from the 
VLA and GMRT, and optical broad-band imaging \citep{cook12,vanzee12} and spectroscopy. 
However, many of the faintest objects are missing the high-quality optical spectroscopy 
needed to determine ``direct" oxygen-abundance metallicity estimates.

As the L-Z relationship provides both a very strong constraint on theories of galaxy 
evolution and a tool to better understand galaxies at higher redshifts \citep{kobulnicky03}, 
we are motivated to better characterize the low-luminosity end of the L-Z relationship. 
Thus, we obtained high-resolution MMT spectroscopy of 42 low luminosity 
star-forming galaxies in the Local Volume with the goal of detecting the [\ion{O}{3}] 
$\lambda$4363 line in order to constrain electron temperature measurements. 

We present our low-luminosity sample in \S~\ref{sec:Sample}, 
with spectral observations obtained from the MMT in \S~\ref{sec:mmt} 
and IRAC photometry in \S~\ref{sec:phot}.
Section~\ref{sec:data} describes the data reduction, followed by the description of
the method used to determine ``direct" oxygen abundances in \S~\ref{sec:metallicity}. 
Our ``Select" sample, compiled from objects with ``direct" oxygen abundances and secure 
distance estimates, is defined in \S~\ref{sec:Gold}.
Using this sample, metallicity is compared to expected trends with \textit{B}-band luminosity, 
4.5 $\mu$m luminosity, and stellar mass in \S~\ref{sec:Lb-Z}, 
\S~\ref{sec:L-Z}, and \S~\ref{sec:M-Z} respectively.
N/O relative abundances are discussed in \S~\ref{sec:N/O}.
In \S~\ref{sec:discussion} we discuss the results of the relationships found in \S~\ref{sec:Lb-Z}-\S~\ref{sec:M-Z},
the ``young galaxy" hypothesis, and the quality of abundance estimators. 
Finally, we summarize our conclusions in \S~\ref{sec:conclusion}.
Appendix~\ref{sec:strong} presents the strong-line abundances for the low-luminosity LVL galaxies for which we were unable to determine ``direct'' abundances.
and Appendix~\ref{sec:outlier} presents our new ``direct" abundances in comparison to the color-temperature metallicity 
relationship of \citet{engelbracht08}.

%----------------------------------------------------------------------------------------------------------------------
%----------------------------------------------------------------------------------------------------------------------

\section{Sample Selection}
\subsection{\textit{Spitzer} LVL Survey}\label{sec:LVL}

LVL is a \textit{Spitzer Space Telescope} legacy program that combines IRAC (Infrared Array Camera) 
and MIPS (Multiband Imaging Photometer) infrared imaging for a complete sample of 258 galaxies 
for the nearest 11 Mpc of our local universe.
These data build upon recent Local Volume galaxy surveys: narrowband H$\alpha$ \citep{kennicutt08}, 
\textit{GALEX} ultraviolet \citep{jlee11}, and \textit{Hubble Space Telescope} resolved stellar population
imaging \citep{dalcanton09}. 
While previous surveys comprehensively cover high surface brightness systems in flux-limited samples,
the LVL survey, although also biased toward high surface brightness galaxies, 
provides a multi-wavelength inventory of a statistically robust, 
approximately volume-limited sample, which is well-suited for studies of dwarf galaxies.
By studying the nearby, low-luminosity galaxies, we can increase the dynamic range
covered by the luminosity-metallicity and mass-metallicity relationships, which will
help to better constrain the slopes.

\subsection{Low-Luminosity LVL Sample}\label{sec:Sample}

We selected a sample of 42 low luminosity galaxies in the LVL survey in order 
to obtain new MMT high-resolution spectra. 
These low luminosity spirals and dwarf irregulars span a range in distance of 2.5 $\leq$ 
D $\leq$ 14.0 Mpc\footnote{Since the inception of the LVL Spitzer program, four galaxies included 
in the sample have updated distances which place them outside of 11 Mpc \citep[see][]{dale09,jlee11}.}. 
The luminosities for this sample range in the near-IR (determined from IRAC \citep{fazio04} photometry)  
from $M_{[4.5]} = -13.1$ to $-21.7$, with \textit{B}-band magnitudes of  $-10.8 \geq M_B \geq -18.8$.
Most of the objects were chosen because they lack ``direct" oxygen abundances in the literature,
their abundance estimates are dated, or were studied with instruments which were known to have problems.

Although not LVL objects, two additional galaxies were added to the sample 
(increasing the sample total to 44 objects) because they played a role in motivating this project.
Both UGC 4393 and UGC 10818 were identified by \citet{engelbracht08}
as low metallicity outliers from the global trend of 70/160 $\mu$m color temperature as a 
function of metallicity.
These two galaxies affect the interpretation of the trend for aromatic emission to 
weaken below 12 + log(O/H) = 7.9 in the mid-IR \citep[see e.g.,][]{engelbracht08} 
and the far-IR \citep[see e.g.,][]{draine07,engelbracht08}.
Because of the possibility that these objects' oxygen abundances were underestimated 
using the lower branch of the R$_{23}$ calibration \citep{pilyugin05}, they were 
included in this sample to be re-examined (see discussion in Appendix~\ref{sec:outlier}).
See Table~\ref{tbl1} for sample characteristics. 

%----------------------------------------------------------------------------------------------------------------------
%----------------------------------------------------------------------------------------------------------------------

\section{DATA}

%----------------------------------------------------------------------------------------------------------------------

\subsection{MMT Spectra}\label{sec:mmt}
\subsubsection{Observations}\label{sec:mmtobs}
New spectroscopy was acquired at the MMT in order to achieve high signal-to-noise 
(S$/$N) spectra with the goal of detecting the faint [\ion{O}{3}] $\lambda$4363 
auroral line at a strength of 4$\sigma$ or higher. 
The observations were obtained with the Blue Channel spectrograph \citep{schmidt89} 
on the UT dates of 2008 October 30-November 1, 2009 June 15-22, and 2010 January 11-12. 
Sky conditions varied, but contained minimal cloud coverage and approximately arcsecond seeing.
A 500 line grating, $1\arcsec$ slit, and UV-36 blocking filter were used, yielding 
an approximate dispersion of 1.2 \AA\ per pixel, a full width at half maximum resolution 
of $\lesssim3$ \AA, and a wavelength coverage of 3690--6790 \AA. 
The sensitivity, resolution, and wavelength coverage of the MMT and Blue Channel spectrograph 
combination allowed for the measurement of all emission lines relevant to oxygen abundance determinations.
Bias frames, flat-field lamp images, and sky flats were taken each night. 
The latter were primarily necessary due to significant differences between the chip illumination 
patterns of the sky and the MMT Top Box that houses the ``BC" incandescent flat-field lamp. 
On average, four standard stars from \citet{oke90} with spectral energy distributions (SEDs) peaking in the 
blue and containing minimal absorption were observed throughout the night using a 
5$\arcsec$ slit over a range of airmasses. 
This allows the flux calibration to be determined as a function of airmass.
The large slit width mitigates the effects of atmospheric differential refraction and allows
accurate measurements of relative fluxes across a large range in wavelength.
Note that since we only care about relative abundances, an absolute flux calibration is not critical.

All 44 galaxies had at least one strong H$\alpha$ brightness peak that was 
aligned with the $1\arcsec\times180\arcsec$ slit. 
Typically, three 900 second exposures\footnote{Some galaxy observations were 
adjusted to shorter or longer exposures depending on the brightness of the 
[\ion{O}{3}] $\lambda$4363 line strength, or included additional exposures 
when the observing program allowed for it; see Table~\ref{tbl2}.} 
were made with the slit at a fixed position angle which approximated the 
parallactic angle at the midpoint of the observation and laid across several 
H$\alpha$ bright regions when possible. 
This, in addition to observing the galaxies at airmasses less than 1.5, served to minimize 
the wavelength-dependent light loss due to differential refraction \citep{filippenko82}. 
A single slit position for each target was deemed sufficient to characterize the 
global oxygen abundance, as metallicity gradients are observed to be small or non-existent in 
low-mass galaxies \citep[e.g.,][]{skillman89, ks96, ks97, lee06b, croxall09}. 
Finally, combined helium, argon, and neon arc lamps were observed at 
each pointing for accurate wavelength calibration. 
A log of the observations is provided in Table~\ref{tbl2}. 
Figure~\ref{fig1} shows the R-band continuum and H$\alpha$ continuum-subtracted 
images for each galaxy, motivating our slit location choices.
The brightest H$\alpha$ regions observed are ordered alphabetically by decreasing flux, 
and the slit positions on the galaxies are shown.  
The images scale as 60x60 arcseconds with North oriented up and East to the left.

%----------------------------------------------------------------------------------------------------------------------

\subsubsection{Spectra Reduction}\label{sec:mmtreduct}
The MMT observations were processed using ISPEC2D \citep{moustakas06}, 
a long-slit spectroscopy data reduction package written in IDL. 
A master bias frame was created from $\gtrsim 20$ zero second exposures by discarding the highest and 
lowest value at each pixel and taking the median. 
Master sky and dome flats were similarly constructed after normalizing the counts in the individual images. 
Those calibration files were then used to bias-subtract, flat-field, and illumination-correct the raw data frames. 
Dark current was measured to be an insignificant $\sim 1$ e$^{-}$ per pixel per hour and was not corrected for. 

Misalignment between the trace of the light in the dispersion direction and the orientation of the CCD detector was 
rectified via the mean trace of the standard stars for each night, providing alignment to within a pixel across the detector. 
A two-dimensional sky subtraction was performed using individually selected sky apertures, 
followed by a wavelength calibration applied from the HeArNe comparison lamps taken at the same telescope pointing. 
Airmass dependent atmospheric extinction and reddening were corrected 
for using the standard Kitt Peak extinction curve \citep{crawford70}. 

For each galaxy, the multiple sub-exposures were combined, eliminating cosmic rays in the process. 
The resulting images were then flux-calibrated using the sensitivity curve derived from 
the standard star observations taken throughout a given night. 
Finally, the trace fit to the strongest continuum source in the slit was used to extract the 
galaxy emission within apertures that encompassed $\gtrsim99\%$ of the light. 
Figure~\ref{fig2} shows a sample of four of the resulting one-dimensional spectra extracted for 
galaxies that had significant [\ion{O}{3}] $\lambda$4363 detections. 
The inset windows display a narrower spectral range to emphasize the [\ion{O}{3}] $\lambda$4363 strength.
This sample does not feature the best spectra from our sample, but rather galaxies are ordered by ionizing 
radiation field strength from highest to lowest as given by the [\ion{O}{3}] $\lambda5007$/[\ion{O}{2}] $\lambda3727$ 
ratio, highlighting the variation within the sample.

%----------------------------------------------------------------------------------------------------------------------

\subsection{Photometry}\label{sec:phot}
To better characterize our low-luminosity sample, absolute magnitudes in 
several different bands were obtained. 
Here we describe their origin and reference their subsequent use.
$M_B$ values were determined by \citet{vanzee12} using 
photometry from apertures matched to the infrared LVL photometry (unless otherwise noted).
Optical photometry for the entire LVL sample is given in \citet{cook12}, whereas \citet{vanzee12} 
focuses on the analysis of colors and EW gradients of dwarf galaxies.
The data are used to examine the optical luminosity-metallicity relationship (see Section~\ref{sec:Lb-Z}).

$M_{[4.5]}$ values from the 4.5 $\mu$m IRAC photometry presented in 
\citet{dale09} were calculated using
\begin{equation}
	M_{[4.5]} = -2.5\log{\frac{F_{[4.5]} (d/10)^2}{179.7}},
\end{equation}
where $F_{[4.5]}$ is the 4.5 $\mu$m flux in Janskys, d is the distance in parsecs, and 179.7 is the zero point flux in Janksys for the 4.5 $\mu$m IRAC band \citep{reach05}.
Distances are taken from the literature, as described in Table~\ref{tbl1}, and assumed to have $10\%$ uncertainty where none were provided.
IRAC calibration uncertainties are $5-10\%$ for the 4.5 $\mu$m data.
Later, in Section~\ref{sec:M-Z}, we use these $M_{[4.5]}$ magnitudes to analyze the NIR luminosity-metallicity relationship. 
Similarly, $M_{K_S}$ values were determined by \citet{dale09} from 2MASS imaging, 
where 666.7 is the zero point flux in Janksys for the 2MASS $K_S$ band.
Although 2MASS $F_{K_S}$ values are available for those objects which \cite{dale09} don't provide $K_S$ magnitudes, 
we choose not to use them. 
The small apertures used in the 2MASS extraction produce unexpectedly faint magnitudes for smaller galaxies 
when compared to similar extractions from IRAC 3.6 and 4.5 $\mu$m data \citep[see, e.g., Figures 4 and 5 in][] {dale09}, and so may not be terribly accurate for our sample.
The $K_S$ magnitudes were used to determine stellar masses in Section~\ref{sec:M-Z}.

Finally, \textit{V}-band magnitudes were needed to calculate $B-V$ colors (see Table~\ref{tbl1}).
When available, $M_{V}$ values were provided by \citet{vanzee12}, using the LVL elliptical aperture.
In other cases, values are taken from \cite{devaucouleurs91} or are determined using \textit{g}- and \textit{r}-band photometry available 
from the Sloan Digital Sky Survey \citep[SDSS;][]{york00}.
The SDSS values are then used to estimate the $B-V$ color following \citet{jester05}:
\begin{equation}
	B-V = \frac{(g-r)+0.22}{1.02}.
\end{equation} 
The available $M_B$, $M_{[4.5]}$, and $B-V$ colors and references for 
this sample are listed in Table~\ref{tbl1}.
Note that the main source of uncertainty in these magnitudes lies in the distance determinations.
Eight of the objects in our sample have distance errors of approximately $10\%$.
Furthermore, 20 of the 44 objects in our sample do not have uncertainties associated with their distance determinations.
For these objects we used an uncertainty of $10\%$, which may be an underestimate for some of them.
The distance uncertainties tend to dominate over the photometric uncertainties. 

%----------------------------------------------------------------------------------------------------------------------
%----------------------------------------------------------------------------------------------------------------------

\section{NEBULAR ABUNDANCE ANALYSIS}\label{sec:data}

%----------------------------------------------------------------------------------------------------------------------

\subsection{Emission Line Measurements}\label{sec:iraf}
Emission line strengths were measured using standard methods available within 
IRAF\footnote{IRAF is distributed by the National Optical Astronomy Observatories, which are operated by the 
Association of Universities for Research in Astronomy, Inc., under cooperative agreement with the National Science Foundation.}. 
In particular, the SPLOT routine was used to analyze the extracted one-dimensional spectra and to fit Gaussian profiles 
to emission lines to determine their integrated fluxes. 
Special attention was paid to the Balmer lines, which are sometimes located in 
troughs of significant underlying stellar absorption. 
The H$\alpha$ emission lines typically had equivalent widths of $\sim$ 350 \AA, 
large enough that the underlying absorption was not a concern.
Even for those H$\alpha$ emission lines with lower EWs, the underlying absorption was negligible.
This was often not the case for H$\beta$ and the lower equivalent width Balmer lines. 
The H$\beta$ absorption EWs for our sample range from 1-8 \AA.
These values are typical of local low-luminosity galaxies, with the majority having H$\beta$ absorption 
EWs between 0 \AA\ and 5 \AA\ \citep[see, e.g., Figure 6 in][]{berg11}.
For the bluer Balmer lines, a multiple component fit was used in which the absorption was fit by a broad, 
negative Lorentzian profile and the emission was fit by a narrow, positive Gaussian profile. 
To ensure a proper fit of the [\ion{O}{3}] $\lambda4363$ line, H$\gamma$ was first fit by a Gaussian
profile, then [\ion{O}{3}] $\lambda4363$ was forced to be fit to the same line profile with the assumption 
that the profile widths of these two neighboring lines should be the same.

Note that we chose to fit the underlying Balmer absorption with Lorentzian profiles, as opposed to using stellar 
population synthesis continuum fitting common in many studies \citep[e.g.,][]{tremonti04}.
Given the large equivalent widths of the Balmer emission lines, the differences between the two methods 
are negligible, and the Lorentzian profiles have the advantage of require no additional assumptions. 
Most importantly, for spectra dominated by young stars, at S/N values typical of our spectra, population synthesis models may not provide a unique solution.
There are also very large variations in the population synthesis models for young ages, 
with large uncertainties in how the Wolfe-Rayet phase, stellar winds, rotation, and other parameters are treated.
Since mass loss and mixing processes in stellar evolution are still poorly understood, 
stellar phases, like Wolf-Rayet stars or Red Super Giants, are particularly affected by such uncertainties \citep{leitherer2011}. 
Later phases, like AGB stars, are covered only crudely in models or not at all, pushing parameters into regimes that are not properly calibrated. 
When discrepancies between models are found, they can usually be attributed to different intrinsic input parameters 
and/or treatment of these aberrant stellar evolutionary phases \citep{vazquez2005,conroy2010}.
By not using the models to fit our continuum, we avoid the uncertainties associated with these implicit assumptions.

The errors of the flux measurements were approximated using 
\begin{equation}
	\sigma_{\lambda} \approx \sqrt{ {(2\times \sqrt{N}\times rms)}^2 + {(0.02\times F_{\lambda})}^2 } ,	\label{eq:uncertainty}
\end{equation}
where N is the number of pixels spanning the Gaussian profile fit to the narrow emission lines. 
The rms noise in the continuum was taken to be the average of the rms 
on each side of an emission line. 
For weak lines, whose uncertainty is dominated by error from the continuum subtraction, 
the rms term determines the approximate uncertainty.
For the lines with flux measurements much stronger than the rms noise of the continuum, 
(usually the H$\alpha$ lines and often the [\ion{O}{3}] $\lambda\lambda$4959,5007 doublet) 
the error is dominated by flux calibration and de-reddening uncertainties. 
In this case, a minimum uncertainty of 2\% was assumed, and the right hand term above dominates the uncertainty estimate.
31 of the 44 galaxies in our sample were measured to have [\ion{O}{3}] 
$\lambda$4363 line strengths $> 4\sigma$.
The measured [\ion{O}{3}] $\lambda4959$/$\lambda5007$ ratios match theoretical expectations within the errors, 
supporting our error estimates and the assumption that the continuum subtraction dominates the uncertainties for 
the weak lines.
For all the objects in the present sample, flux line strengths and corresponding 
errors are listed in Table~\ref{tbl3}. 
We concentrate the rest of our analysis on the objects for which direct electron 
temperature and chemical abundance determinations can be made.
An analysis of the remaining spectra using strong-line methods is reported in Appendix~\ref{sec:strong}.

%----------------------------------------------------------------------------------------------------------------------

\subsection{Reddening Corrections}\label{sec:redcor}
The relative intensities of the Balmer lines are nearly independent of both 
density and temperature, so they can be used to solve for the reddening. 
The MMT spectra were de-reddened using the reddening law of \citet{cardelli89}, 
parameterized by $A_{V}=3.1\ E(B-V)$, where the extinction, $A_{1}(\lambda)$ 
was calculated using the York Extinction Solver 
\citep{mccall04}\footnote{http://www1.cadc-ccda.hia-iha.nrc-cnrc.gc.ca/community/YorkExtinctionSolver/}. 
With these values, the reddening, $E(B-V)$, can be derived using
\begin{equation}
	\log{\frac{I(H\alpha)}{I(H\beta)}\ } = \log{\frac{F(H\alpha)}{F(H\beta)}\ } + 0.4\ E(B-V)\ [A_{1}(H\alpha)-A_{1}(H\beta)],
	\label{eq:dered} 
\end{equation}
where $F$(H$\alpha)/F$(H$\beta$) is the observed flux ratio and $I$(H$\alpha)/I$(H$\beta$) 
is the de-reddened line intensity ratio using case B from \citet{storey87}, assuming 
an electron temperature calculated from the [\ion{O}{3}] line ratio and $n_{\rm e}=10^{2}$ cm$^{-3}$.
For our sample, the electron temperature range is 9,500 K - 19,500 K, with an average of 13,300 K.
This range agrees with the typical electron temperatures of 10,000 K - 20,000 K for metal-poor \ion{H}{2} regions.   
This same process can be carried out for the H$\gamma$/H$\beta$ and H$\delta$/H$\beta$ ratios observed.
When all the necessary Balmer lines were present, which is true of all of the objects in our ``Select" sample,
we used a minimized chi squared approach to find the best estimate of E(B-V) based on the H$\alpha$/H$\beta$, H$\gamma$/H$\beta$, and H$\delta$/H$\beta$ ratios.
The resulting Balmer ratios are within errors of the \citet{storey87} Case B values for all objects meeting the 
selection criteria of our ``Select" sample (see \S~\ref{sec:Gold}), with an average of $\chi^2=0.03$.

Following \citet{lee04}, the reddening value can be converted to the logarithmic extinction at H$\beta$ as
\begin{equation}
	c(\mbox{H}\beta) = 1.43\ E(B-V).
	\label{eq:cHbeta}
\end{equation}
Our reddening corrections are tabulated in Table~\ref{tbl3}.

%----------------------------------------------------------------------------------------------------------------------
%----------------------------------------------------------------------------------------------------------------------

\section{``Direct" OXYGEN ABUNDANCE DETERMINATIONS}\label{sec:metallicity}

Accurate ``direct" oxygen abundance determinations from \ion{H}{2} regions require 
a measurement of the electron temperature (typically via observation of the 
temperature sensitive auroral [\ion{O}{3}] $\lambda$4363 line).  
For the 31 low-luminosity objects for which [\ion{O}{3}] $\lambda$4363 strengths were 
measured to be $> 4\sigma$, we use the temperature sensitive ratio comparing 
``auroral" to ``nebular" collisionally excited lines to determine electron temperatures.
A simple, yet reasonable, approximation to the geometry of an 
\ion{H}{2} region is to assume a two zone volume, where $t_2$ and $t_3$ are the electron 
temperatures (in units of $10^{4}$ K) in the low and high ionization zones respectively.
For the high ionization zone, the [\ion{O}{3}] I($\lambda\lambda$4959,5007)/I($\lambda$4363) 
ratio was used to derive a temperature using the IRAF task TEMDEN. 
This task computes the electron temperature of the ionized nebular gas within the 5-level atom approximation.
The O$^{+}$ (low ionization) zone electron temperature can be related to the O$^{++}$ 
(high ionization) zone electron temperature \citep[e.g.,][]{campbell86,pagel92}.
We used the relation between $t_{2}$ and $t_{3}$ proposed by \citet{pagel92}, 
based on the photoionization modeling of \citet{stasinska90} to determine 
the low ionization zone temperature:
\begin{equation}
	{t_{2}}^{-1} = 0.5({t_{3}}^{-1} + 0.8).
\end{equation}
The low and high ionization region temperatures are tabulated in Table~\ref{tbl4}. 
Typically \ion{H}{2} regions are assumed to have electron temperatures within the range of 1 to 2 $\times10^4$ K.
Temperatures for the present sample agree with this approximation, spanning 10,800 K - 15,200 K 
for the low ionization region, and 9,600 K - 19,400 K in the high ionization region.

Since the MMT spectra include emission lines from both O$^+$ and O$^{++}$, 
we determine oxygen abundances based on our estimated two zone electron temperatures.  
Spectra which contained measurable [\ion{S}{2}] $\lambda\lambda$6717,6731 were used to 
determine electron densities consistent with the low density limit. 
Thus, it is reasonable to simply assume $n_{\rm e}=10^2$ cm$^{-3}$ for this sample.
Ionic abundances were calculated with:
\begin{equation}
	{\frac{N(X^{i})}{N(H^{+})}\ } = {\frac{I_{\lambda(i)}}{I_{H\beta}}\ } {\frac{j_{H\beta}}{j_{\lambda(i)}}\ }.
	\label{eq:Nfrac}
\end{equation}
The emissivity coefficients, which are functions of both temperature and density, were determined using 
the IONIC routine in IRAF. 
This routine applies the 5-level atom approximation, assuming the appropriate 
ionization zone electron temperature, as determined from the oxygen line ratios.

Some abundance determinations require ionization correction factors 
to account for unobserved ionic species. 
Here we assume N/O = N$^{+}$/O$^{+}$ \citep{peimbert69}.
\cite{nava06} have investigated the validity of this assumption.  
They concluded that although it could be improved upon with modern photoionization models, 
it is valid to within about 10\%.
Thus, we employ this assumption, mostly for the purposes of direct comparison with other 
studies in the literature.

For the 9 objects with multiple \ion{H}{2} regions containing strong 
[\ion{O}{3}] $\lambda$4363, an error weighted average was used to determine 
a best estimate of relative abundances and oxygen abundances.
The results from individual \ion{H}{2} regions are tabulated in 
Table~\ref{tbl4} and the mean values, using a weight of $1/\sigma_i^2$ for each component, are listed in Table~\ref{tbl5}.
The uncertainties for these mean values are represented by the standard deviation of the weighted mean or the weighted dispersion, which ever is greater.
Calculated errors in this paper provide a statistical estimate only.
Additional errors may be important, such as systematic errors due to temperature fluctuations or other imperfect assumptions.
However, the purpose of this paper is to improve the L-Z and M-Z relationships with abundances from high quality spectra.
The statistical errors allow such an assessment of the relative quality of the spectra used, which in turn are weighted higher in the regression fits.

For 7 of the 9 dwarf galaxies with direct abundances from multiple \ion{H}{2} regions,
the derived oxygen abundances agree within the uncertainties.
These support the interpretation that the ISM in typical dwarf galaxies is chemically well mixed,
in agreement with past studies \citep[e.g.,][]{skillman89, ks96, ks97, lee06b, kehrig08, croxall09, perez-montero2011}. 
Various theoretical studies support this result \citep[e.g.,][]{roy95}.
However, there are two galaxies for which the oxygen abundances don't agree. 
For NGC 4449 the highest signal to noise spectrum is offset to higher log(O/H) values by 0.16 and 0.18 dex compared to the other two.
This discrepancy may be due to the possible contamination of an embedded supernova remnant \citep[e.g.,][]{skillman85}, or it may be truly offset.  
Additional spectra are needed to clarify this.
NGC 2537 has two high quality optical spectra, but the derived values disagree by 0.26 dex.  
This factor of nearly two difference is intriguing, warranting further investigation of this object.  
We increased the error of the weighted mean to indicate the dispersion between the two values.  
Note that the lower value would be in better agreement with the L-Z relationships, but that the mean is not offset very far.
Overall, the oxygen abundances determined in this paper are all relatively low 
(12 + log(O/H) $< 8.3$; average 12 + log(O/H) $= 7.84$) as we would expect for low-mass, low-luminosity galaxies.
The abundances for the two additional objects outside of the LVL sample, UGC 4393 and UGC 10818, 
are discussed in Appendix~\ref{sec:outlier}.

%----------------------------------------------------------------------------------------------------------------------
%----------------------------------------------------------------------------------------------------------------------

\section{The L-Z and M-Z Relationships}\label{sec:relationships}

The new ``direct" oxygen abundances determined in this paper provide an 
opportunity  to expand relationships previously limited by the reliability of empirical calibrations.  
In particular, these measurements allow us to re-examine the L-Z and M-Z relationships derived by L06, 
which are limited by small number statistics at the low luminosity end.

%----------------------------------------------------------------------------------------------------------------------

\subsection{The Total and ``Select" Samples}\label{sec:Gold}

In the following, we analyze various samples based on both abundance 
measurement and distance measurement quality criteria.
Specifically, we label the samples of galaxies with both direct oxygen 
abundance measurements and accurate distances as ``Select.''
We observed 31 objects with [\ion{O}{3}] $\lambda4363$ detected at 
a strength greater than 4$\sigma$; this comprises our total sample.
Our ``direct" oxygen abundance measurements have relatively small errors, but comparisons 
to luminosity and stellar mass calculations require accurate distance determinations.  
This motivated further cuts from our sample to keep only objects with reliable distance 
determinations using the tip of the red giant branch (TRGB) or Cepheid variables (ceph), 
giving rise to our 13 object ``Select" sample. 
In addition, the L06 data were updated with 4.5 $\mu$m photometry from \citet{dale09} 
(to minimize the effects of aperture differences between the previous photometry and our own), 
distances from \citet{dalcanton09}, and ``direct" oxygen abundances from \citet{croxall09} when available.
Those objects that passed the selection criteria were assembled into a similar ``Select L06" sample of 14 objects.
Other Local Volume objects presented in \citet{vanzee06a} and \citet{marble10} were considered for an additional ``Select" sample.
Using the same criteria mentioned above, this  provided 11 additional objects with 
``direct" abundances at a strength of $4\sigma$ or greater and accurate TRGB distances. 
The 13 ``Select" objects from this paper are noted in Table~\ref{tbl5} and the properties of the additional
objects taken from the literature are listed in Table~\ref{tbl6}.
Together these data sets made the final ``Combined Select" sample comprised of 38 objects with both 
secure distance (TRGB or ceph) and oxygen abundance determinations ([\ion{O}{3}] $\lambda 4363 > 4\sigma$).
Note that we have 18 objects with accurate oxygen abundances that require accurate distances from TRGB 
observations in order to be elevated to the ``Select" caliber.  Of these, 13 have distances less than
8 Mpc, so their TRGB distances could be obtained with a relatively small investment of \textit{Hubble Space Telescope} time.

Due to the wealth of \textit{B}-band photometry available from previous studies, 
the majority of the sample has \textit{B}-band absolute magnitude estimates.
With the addition of \textit{Spitzer} IRAC photometry, all members of the ``Select" 
sample also have 4.5 $\mu$m absolute magnitudes as determined by \citet{dale09}.
In the following sections we discuss the low-luminosity portion of both the optical and NIR L-Z relationships 
and the subsequently determined M$_{\star}$-Z relationship, for our whole 
sample of ``direct" oxygen abundances and a comparison to the filtered ``Combined Select" sample.

%----------------------------------------------------------------------------------------------------------------------

\subsection{\textit{B}-band L-Z Relationship}\label{sec:Lb-Z}

In the top panel of Figure~\ref{fig:LbZ} we compare ``direct" metallicities to corresponding \textit{B}-band luminosities.
Taking into consideration the errors on both quantities \citep[c.f.,][]{press92}, we determine the most likely linear
fit to the data using the MPFITEXY routine \citep{williams10}, which depends, in turn, on the MPFIT package \citep{markwardt09}.
In this section, and those following, we provide the total scatter (intrinsic $+$ observational) output from the MPFITEXY routine, which is essentially
a weighted mean of the scatter of the data about the linear fit.
In each case, we compare our results to that of L06, who also use a weighted dispersion routine.

The best fit to the 31 objects in the current sample with ``direct" oxygen abundance 
measurements results in:
\begin{equation}
	12 + \log(\mbox{O/H}) = (6.59\pm0.32) + (-0.08\pm0.03) M{_B},
\end{equation}
with a dispersion in log(O/H) of $\sigma=0.19$. 
Updated data for the L06 sample (see \S~\ref{sec:Gold}) is also plotted, and compared to the original least-squares best fit of L06.

The low-metallicity outlier at 12 + log(O/H) = 7.20 is the blue compact dwarf UGC 5340, supporting its classification 
by previous work as one of the most metal-deficient star-forming galaxies \citep[e.g.,][]{izotov07, pustilnik08b}.
However, \citet{pustilnik08b} note that its present distance could be significantly \textit{underestimated} due to the large 
negative peculiar velocity in that region, which, if true, would result in an even larger discrepancy. 
\citet{ekta08} and \citet{pustilnik08a} have discussed the HI observations of UGC 5340 and concluded that it 
is likely undergoing a merger, which could explain, at least in part, its discrepant position from the L-Z relationship.  
From HI observations of a sample of extremely metal poor galaxies, \citet{ekta10} find that roughly half of 
these galaxies show evidence of interactions, and conclude that the very low metallicities in these galaxies 
are due to recent infall of metal poor gas \citep[see also][]{jlee2004}).  
Thus, these galaxies do not lie on the L-Z relationship defined by the average low luminosity galaxy, and therefore,
UGC 5340 has not been included in the relationships of the ``Combined Select" 
sample.\footnote{At this time the \ion{H}{1} morphologies have not been analyzed for the LVL sample, so we cannot make 
predictions about the infall of unenriched gas for these galaxies.}

In the lower panel of Figure~\ref{fig:LbZ} we plot the 38 objects in the ``Combined Select" sample. 
The best fit is given by:
\begin{equation}
	12 + \log(\mbox{O/H}) = (6.27\pm0.21) + (-0.11\pm0.01) M{_B}.
\end{equation}
with a resultant dispersion in log(O/H) of $\sigma=0.15$\footnote{Dispersion in log(O/H) of 
the ``Combined Select" sample increases to $\sigma=0.18$ if UGC 5340 is included.}. 
Note that the luminosity error bars represent the error propagated from the 
uncertainty in the photometry and distances.
This relationship agrees with that of L06 within errors.
Additionally, the MPFITEXY routine allows us to estimate the intrinsic scatter by
ensuring that $\chi ^2$/(degrees or freedom) $\approx$ 1.  
Using this tool, the intrinsic scatter in log(O/H) for the \textit{B}-band L-Z relationship for 
the ``Combined Select" sample is 0.13 dex, i.e., most of the scatter in this relationship is intrinsic. 

%----------------------------------------------------------------------------------------------------------------------

\subsection{4.5 $\mu$m L-Z Relationship}\label{sec:L-Z}

L06 found their L-Z slope to be smaller in the NIR than in the optical and to contain less scatter.
This result might be expected since luminosities in redder bands are less sensitive to 
dust extinction and star formation rates than optical luminosities.
However, these NIR luminosities are also vulnerable to stochastic effects from the high NIR luminosities of AGB stars.
Following the motivation given in L06, we analyze the 4.5 $\mu$m L-Z relationship.

In the top panel of Figure~\ref{fig:LZ}, we plot the 4.5 $\mu$m L-Z relationship
for our low-luminosity LVL sample.
Our results are well matched to the luminosity-metallicity relationship 
for dwarf galaxies found by L06 \citep[and corroborated by][]{marble10}. 
Using the MPFITEXY least-squares fit to our data, the resulting expression is:
\begin{equation} 
	12+\log(\mbox{O/H})= (6.37\pm0.33) + (-0.08\pm0.02) M_{[4.5]},
\end{equation}
with a standard deviation in log(O/H) of $\sigma=0.18$.
The original L06 least-squares fit and the updated L06 data are also plotted in Figure~\ref{fig:LZ},
displaying an equivalent slope, but with a notably smaller dispersion in log(O/H) of only 0.12. 
Note that while the two fits have the same slope, they are offset from one another by roughly 0.1 dex in log(O/H); 
this difference is within the error and can be attributed to the difference in samples and small sample size.

In the bottom panel of Figure~\ref{fig:LZ} we have plotted the NIR L-Z 
relationship for the ``Combined Select" sample.
A least-squares fit results in:
\begin{equation} 
	12+\log(\mbox{O/H})= (6.10\pm0.21) + (-0.10\pm0.01) M_{[4.5]}
\end{equation}
and produces a standard deviation of $\sigma=0.14$\footnote{Dispersion in log(O/H) of 
the ``Combined Select" sample increases to $\sigma=0.22$ if UGC 5340 is included.}.
This is nearly identical to the standard deviation of $\sigma=0.15$ found for the ``Combined Select" 
sample for the optical L-Z relationship, and the slopes are the same within the uncertainties. 

The intrinsic scatter in log(O/H) for the 4.5 $\mu$m L-Z relationship for
the ``Combined Select" sample is 0.11 dex.  
Since AGB stars can have significant impact on the NIR luminosities, we must consider the effect of 
stochastic sampling on the overall scatter of our relationship.
However, since we find such a small scatter in the NIR L-Z relationship it is unlikely to be due to AGB stars, 
which would normally drive the data to a larger dispersion.
L06 determined dispersions in the optical and NIR L-Z relationships of 0.161 and 0.122 respectively.
In comparison, the present work does not find a significant difference between the dispersions of the NIR 
and optical L-Z relationships.
However, the NIR intrinsic scatter in log(O/H) is slightly smaller than the intrinsic scatter for the 
\textit{B}-band L-Z relationship for the ``Combined Select" sample (0.11 versus 0.15 dex). 

%----------------------------------------------------------------------------------------------------------------------

\subsection{M$_{\star}$-Z Relationship}\label{sec:M-Z}
The underlying relationship between mass and luminosity and the relative ease of 
measuring luminosities has allowed a widespread use of the L-Z relationship.
However, mass is thought to be more fundamentally related to metallicity \citep[see, e.g.][]{tremonti04}, 
and so, when possible, metallicity is also investigated as a function of stellar mass. 
In order to examine the M$_{\star}$-Z relationship, we need to estimate stellar masses in a self consistent way.
Although SED fitting is commonly used to determine individual masses,
the necessary spectral and/or photometric components were not available to us for our entire ``Combined Select" sample.
Stellar mass can also be inferred from luminosity, where optical colors have 
been widely used to estimate M/L ratios \citep[e.g.,][]{brinchmann00,bell01}.
It is important to note the uncertainties in M/L ratios that occur due to variations in the current star formation rate,
which are most significant if galaxies have formed a substantial fraction ($>$10\%) of their stars in a recent episode.
Near IR magnitudes are often a better choice to characterize the galaxy luminosity because they 
are less sensitive than bluer bands to extinction and the age of the stellar population.
The dominant emission in NIR wavelengths arises from the stellar populations (as opposed to dust) and is only marginally sensitive to recent star formation, but even so, NIR stellar M/L ratios can vary by up to a factor of 
$\sim$2 due to the star formation rate and stellar metallicity \citep{bell01}. 
Furthermore, \citet{lee06a} found that although individual stellar masses can vary by as much as 
$\sim$0.5 dex with M/L model, the subsequent M-Z relationship spanning four decades in stellar mass is nearly 
independent of the model chosen. 

We chose to estimate stellar mass in a uniform manner from 4.5 $\mu$m luminosity and $K-[4.5]$ and 
$B-K$ color following the method presented by L06:
\begin{equation}
	\log{\mbox{M}_{\star}}=\log(\mbox{M}_{\star}/L_{K}) + [\log{L_{[4.5]}} - 0.4\ (K-[4.5])].
	\label{eq:Mstar}
\end{equation}
L06 derived a mass-to-light ratio ($\mbox{M}_{\star}/L_{K}$) as a linear function of 
\textit{B$-$K} color based on the Bruzual \& Charlot model with a Salpeter IMF.
Note that there is a systematic uncertainty in NIR M/L ratios of $\sim$0.2 dex due to uncertainties in AGB evolution \citep[e.g.,][]{conroy2010,melbourne12}.
Since $K_s$ photometry is available for the LVL sample \citep{dale09}, 
unlike the procedure of L06, the $B-K$ color was calculated directly 
(we assume M$_K$ $\equiv$ M$_{K_{s}}$).
Based on the direct relationship between the ratio of luminosities and ratio of 
absolute magnitudes for two objects, we calculated monochromatic luminosities, 
$L_{[4.5]}$, assuming $M_{[4.5]}\simeq3.3$ for the Sun (following the logic of L06).
The M$_{\star}$ results are tabulated in Table~\ref{tbl1}.

In principle, mass estimates can be improved using SED fitting to broad-band photometry 
which span from the UV to the IR. 
\citet{johnson12} have determined masses for the LVL galaxies using this method.  
Unfortunately, the broad wavelength coverage and associated analysis is not available 
for the entire LVL survey, including objects in our sample. 
There are 41 LVL galaxies for which we have obtained new spectra or which have spectra 
in the literature with masses computed by \citeauthor{johnson12} to which we can compare
our stellar masses determined from 4.5$\mu$m luminosities.
We find an average difference of 0.23 dex in mass, or an offset of a factor of $\sim$2, 
in the sense that the SED derived masses are smaller and independent of luminosity or optical color.  
This difference can be accounted for by the use of different IMFs in the modeling (Salpeter IMF in \cite{bell01} 
and Chabrier IMF in \cite{johnson12}).
Note that this average difference, as well as the dispersion of $\sigma$=0.24, 
is smaller than the typical uncertainty in our derived masses.
Therefore, adopting these masses would not affect the slope of our derived M-Z relationship. 
Because we do not have SED derived masses for our entire ``Combined Select" sample, 
we report the present relationship using the masses calculated here.

M$_{\star}$-Z data are plotted in the top panel of Figure~\ref{fig:MZ} in comparison 
to the updated L06 data and original M$_{\star}$-Z relationship of L06. 
The best fit to our data, 
\begin{equation} 
	12+\log(\mbox{O/H})= (5.43\pm0.42) + (0.30\pm0.05) \log(\mbox{M}_{\star}),
\end{equation}
with a dispersion of $\sigma=0.21$, agrees, within errors, with the fit to the L06 data set.
This dispersion is notably larger than the 0.12 dispersion in log(O/H) found by L06.
The mass error bars used here are the propagated errors from the 4.5 $\mu$m luminosity, K-[4.5] color,
and mass-to-light ratio (where we substituted the uncertainty in B-K color).
Note that the contrast in dispersion of the two data sets is largely due to the different errors.
L06 assumed the same errors for their mass determinations as their 4.5 $\mu$m luminosities,
whereas we incorporated the additional propagated error from the color terms.
This difference accounts for the disparity in uncertainty.

On the bottom of Figure~\ref{fig:MZ} we have plotted the ``Combined Select" M-Z data.
Fitting the combined data set produces the least-squares linear fit, 
\begin{equation} 
	12+\log(\mbox{O/H})= (5.61\pm0.24) + (0.29\pm0.03) \log(\mbox{M}_{\star}),
	\label{eqMZ}
\end{equation}
with a standard deviation of $\sigma=0.15$\footnote{Dispersion in log(O/H) of 
the ``Combined Select" sample increases to $\sigma=0.21$ if UGC 5340 is included.}, 
which is essentially equivalent to the dispersions of the ``Combined Select" L-Z data sets.
The intrinsic scatter in log(O/H) for the M$_{\star}$-Z relationship for the ``Combined Select" sample is 0.08 dex.  
This appears to be significantly smaller than the intrinsic scatter in log(O/H) for the 4.5 $\mu$m L-Z 
relationship for the ``Combined Select" sample of 0.11 dex.

The dual effects of increasing the number of objects observed and selecting only objects
with both reliable oxygen abundances and distances has resulted in a better characterization
of the L-Z and M-Z relationships.  
In this work, we assume that a galaxy with an \ion{H}{2} region of sufficiently high surface brightness 
to allow a $\lambda$4363 measurement is a local property of the star forming region, 
and not related to a characteristic property of the host galaxy.  
Thus, we don't believe our sample to be biased in terms of mass or galaxy type.  
Additionally, the observation that strong-line abundances of low-mass galaxies are consistent with 
the relationships derived here, albeit with increased scatter, supports this assumption.
Therefore, the L-Z and M-Z relationships presented here should accurately represent low-mass galaxies in general.
In high mass galaxies, \citet{tremonti04} found a decrease in the dispersion in the L-Z relationships as 
one went from $\sigma=0.16$ for the optical B-band to $\sigma=0.13$ for the longer wavelength z-band, 
and then an even smaller dispersion of $\sigma=0.10$ for the M-Z relationship.
The ``Combined Select" data show a negligibly smaller dispersion for the NIR L-Z relationship compared to the 
B-band, and no similar decrease in dispersion for the M-Z relationship.

%----------------------------------------------------------------------------------------------------------------------

\section{N/O Relative Abundances}
\label{sec:N/O}
The N/O versus O/H trend is well studied in galaxies of varying types.
\citet{vce93} presented a thorough overview of theoretical expectations 
and observations available at the time.
A salient point is that N can be produced as both a primary and a secondary 
element and that the secondary component is expected to be delayed 
relative to oxygen and to dominate at high abundances.
A typical scenario might be described by oxygen production in Type II supernovae 
being released 10 Myr after star formation, whereas nitrogen forming in intermediate 
mass stars isn't released until much later times \citep[$> 10^{8}$ Myr;][]{ks96}.
Initially, N/O is expected to rapidly decrease as oxygen is returned to the interstellar medium,
but will gradually increase with time as nitrogen begins to be returned to the gas reservoir.
Thus, in principle, the relative N/O abundance can be used as a clock \citep[e.g.,][]{henry00} to 
indicate the time since the most recent burst of star formation. 
Note that this effect is not expected if the star formation rate does not show significant
variations \citep{molla06}.

Table~\ref{tbl5} lists the error weighted average N/O values for our sample.
The N/O errors were determined by first adding in quadrature the error in flux of 
both [\ion{O}{2}] $\lambda3727$ and [\ion{N}{2}] $\lambda6584$, then adding 
this value in quadrature with the error in temperature of the low ionization zone.
The most extreme values extend from log(N/O) = $-1.77$ to $-1.00$, with an 
average of log(N/O) = $-1.47$; this is comparable to the isolated dwarf irregular 
sample examined by \citet[][hereafter vZ06]{vanzee06b}, with an average log(N/O) = $-1.41$.
We tested for a correlation of N/O with reddening and found none, indicating an 
absence of bias in this regard. 

The 9 objects with multiple ``direct" oxygen abundances provides the 
opportunity to study N/O variations in individual dwarf galaxies.
The average N/O ratio dispersion of different \ion{H}{2} regions in a given galaxy is only 0.08 dex, 
indicating that dwarf galaxies, despite appearing to be solid body rotators \citep{skillman88}, 
are well mixed \citep[see also e.g.,][]{roy95}.
Other studies, such as the \textit{green pea} galaxies analyzed by \citet{amorin10} and the nitrogen 
enriched dwarf galaxies analyzed by \citet{perez-montero2011}, find N/O abundance dispersions 
or small gradients hypothesized to be a combination of outflows of enriched gas and inflows of metal-poor gas.
Note that errors in N/O account for the dispersion within four of the objects that have multiple 
N/O measurements (UGC 1056, UGC 4278, NGC 3738, and NGC 4449),
but not for 5 others (NGC 784, NGC 2537, UGC 4393, UGC 5423, and UGC 8638).
For two of these objects (NGC 784 and UGC 4393) the differences in N/O are significant (0.19 and 0.15).  
In these last two cases in could be that significant nitrogen enhancement has been detected, 
although not at the level of the well studied galaxy NGC~5253 \citep[e.g.,][]{ksrwr97,ls12} 
or the more recently discovered N/O anomaly in MRK~996 \citep{james09}.

vZ06 looked at several variables for their possible influence on N/O abundance.
In particular, they found a correlation between N/O and color, in the sense that redder 
galaxies have higher N/O as one might expect from time delayed N release.
In the top panel of Figure~\ref{fig:NO}, log(N/O) is plotted vs. $B-V$ color for objects of our sample with 
``direct" abundances and measurable [\ion{N}{2}]/[\ion{O}{2}] abundances. 
Similar to vZ06, we find a fairly steep increase in N/O with redder color
(demonstrated by the dotted least squares fit):
\begin{equation}
        \log(\mbox{N/O})=  (-1.96\pm0.12) + (1.22\pm0.26)\times(B-V).  
\end{equation}
with a dispersion of $\sigma$ = 0.13.
In fact, the two groupings of points are visually consistent with one another.
When the additional objects from the literature are added to the plot, 
the least squares fit over 0.05 $\lesssim B-V \lesssim$ 0.75 to all of the data is
\begin{equation}
        \log(\mbox{N/O})=  (-1.92\pm0.08) + (1.18\pm0.19)\times(B-V),  
\end{equation}
which agrees well with the relationship found by vZ06.
Below $B-V$ = 0.10 there are two objects with discrepantly large N/O values.
Therefore, we suggest this fit is most appropriate for the range of 0.20 $\lesssim B-V \lesssim$ 0.75.
Note the appearance of significant scatter in this figure.
We calculate a dispersion in log(N/O) of $\sigma$ = 0.14 dex, with an estimated intrinsic scatter 
of 0.10 dex.

Additionally, the bottom panel of Figure~\ref{fig:NO} shows log(N/O) plotted vs.\ 12 + log(O/H) 
for the same sample.
Above 12 + log(O/H) $\approx$ 7.7 a trend of N/O increasing with O/H is evident, despite the large scatter.
For 12 + log(O/H) $\ge$ 7.7, the best fit to our data yields:
\begin{equation}
        \log(\mbox{N/O})=  (-5.49\pm1.36) + (0.51\pm0.17)\times[12+\log(\mbox{O/H})],  
\end{equation}
with a dispersion of $\sigma$ = 0.16, where the estimated intrinsic scatter is 0.14 dex. 
With an increasing slope, this would be indicative of secondary N production in this region.
\citet{garnett90} proposed that much of the scatter in the 12 + log(O/H) vs 
log(N/O) relationship could be explained by the time delay between 
producing oxygen and secondary nitrogen.

For the systems with 12 + log(O/H) $\le$ 7.7, in agreement with previous studies, there is little trend in N/O
with O/H.  We have calculated a weighted mean in N/O using the IDL routine MPFITEXY with the added constraint
of setting the slope to zero for the points below 12 + log(O/H) $=$ 7.7.
For our eight new observations, the weighted mean is log(N/O) = $-$1.56 with a standard deviation of 0.05. 
For the nine observations from the literature, the weighted mean is log(N/O) = $-$1.51 with a standard deviation of 0.04.
For the two sets together we obtain log(N/O) = $-$1.56 with a standard deviation of 0.05.
Of this dispersion, the intrinsic scatter is predicted to be 0.02, so observational scatter may play a large role 
in determining the observed scatter in this relationship.
In most previous studies, no correlation is noted between 12 + log(O/H) and the relative N/O abundance at low 
oxygen abundances, where nitrogen is expected to behave like a primary nucleosynthesis element.
Together the new observations are consistent with the trends in N/O with O/H observed by 
\citet{vce93}, \citet{jlee2004}, \citet{vanzee06a}, \citet{molla06}, and \citet{liang06}.

%----------------------------------------------------------------------------------------------------------------------
%----------------------------------------------------------------------------------------------------------------------

\section{Discussion}
\label{sec:discussion}

\subsection{The L-Z and M-Z Relations for Low Luminosity Galaxies}
\label{sec:low}

The dual effects of increasing the sample size and selecting only objects with both reliable oxygen 
abundances and distances has resulted in an improved characterization of the L-Z and M-Z relationships.
In high mass galaxies, \citet{tremonti04} found a decrease in the dispersion in the L-Z relationship as one 
went from the optical B-band ($\sigma=0.16$) to the longer wavelength z-band ($\sigma=0.13$), and an even 
smaller dispersion for the M-Z relationship ($\sigma=0.10$).
The present data show only a slightly smaller dispersion for the NIR L-Z relationship ($\sigma=0.14$) 
compared to the B-band ($\sigma=0.15$), but no similar decrease in dispersion for the 
M-Z relationship ($\sigma=0.15$).  
However, our estimates of the \textit{intrinsic} scatter in the three relationships do show a decreasing trend in the 
sense that the \textit{intrinsic} scatter of the B-band L-Z relationship is largest ($\sigma=0.13$),
followed by the NIR L-Z relationship ($\sigma=0.12$), then the M-Z relationship ($\sigma=0.08$).  
While this trend could be an artifact of how the errors are estimated for the three different parameters,
it is interesting that it follows the same pattern observed in the larger spiral galaxies.
Perhaps what is most remarkable is the small \textit{intrinsic} scatter in all three relationships.
When averaging the light over an entire galaxy, as done in \citet{tremonti04} one might 
expect relatively low dispersions.  
However, oxygen abundances derived from spectroscopic apertures only covering a fraction of 
the galaxy will be biased if radial gradients exist \citep[e.g.,][]{moustakas12}.
Therefore, one might expect much larger dispersions when observing individual \ion{H}{2} regions, 
yet this is not the case observed in most dwarf galaxies, as they have been shown to be relatively 
chemically homogeneous \citep[e.g.,][]{croxall09}. 
	
The L-Z and M-Z relationship slopes determined for the ``Combined Select" sample are similar to 
those found in previous studies \citep[e.g.,][]{tremonti04,lee06a}.
For large galaxies, a different slope may apply as galaxies higher in mass and luminosity contain 
more metals and dust \citep[e.g.,][]{rosenberg06} causing them to appear under-luminous.
For smaller, less luminous galaxies, even with the present sample included, 
the number of galaxies meeting our ``Select" criteria is still relatively small.
This limitation could affect our measurements of the scatter, but it appears
that these relationships have intrinsically smaller dispersions.
The evolutionary paths of dwarfs are still poorly understood, making the source of this inherent variation unclear.
Some studies argue for the importance of gas infall and outflows \citep[e.g.,][]{garnett02}, whereas others point to
star formation efficiencies \citep[e.g.,][]{lequeux79,brooks07}, and variations in initial mass functions \citep[e.g.,][]{koppen07}.
Still other studies have also seen significant scatter at low stellar masses \citep[see for example][]{tremonti04,amorin10}.

\cite{amorin10} suggest that inherent variation in the L-Z and M-Z relations could result from these 
objects being relatively young and thus may still be converting large amounts of cold gas into stars. 
If these young galaxies have not had enough time for several generations of star formation to produce
massive AGB stars, then we would expect very little absorption due to dust.
The relative uniformity between the dispersions of the L-Z and M-Z relationships and between the slopes of 
the optical and near-IR L-Z relationships is consistent with this idea, suggesting no more absorption in the 
optical than in the near-IR, and thus very little dust is present in these low-luminosity galaxies.
The fact that the scatter in the L-Z and M-Z relationships is small suggests that AGB stars do not play as 
significant of a role in determining the scatter in the NIR L-Z relationship for low-mass galaxies.
In fact, in our sample it seems that AGB stars are balanced out by the effects of star formation histories. 
Whatever the actual source of the scatter may be, since we used the most reliable 
oxygen abundances and distance estimates possible in constructing the L-Z and M-Z relationships, 
it appears that the dispersion for this sample is real as it is larger than observational errors.
However, the ``young galaxy'' hypothesis faces other observational challenges.

%----------------------------------------------------------------------------------------------------------------------

\subsection{N/O and the Young Galaxy Hypothesis}
\label{sec:young}

\citet{garnett90} first showed that the N/O ratio in low metallicity star forming galaxies is relatively constant 
as a function of O/H (with a mean value of log(N/O) = $-$1.46$^{+0.10}_{-0.13}$) for these ``plateau'' objects.
Later, \citet{izotov99} drew attention to the plateau with small dispersion in log (N/O) ($-$1.60 $\pm$ 0.02) 
in extremely metal-poor (12 + log(O/H) $\le$ 7.6) blue compact dwarf galaxies.   
They proposed that the absence of time-delayed production of N (and C) is consistent with the scenario that 
extremely metal-poor galaxies are now undergoing their first burst of star formation, 
and that they are therefore young, with ages not exceeding 40 Myr.  
They further argued that if this were true, then this would argue against the commonly held belief that C and N 
are produced by intermediate-mass stars at very low metallicities (as these stars would not have yet completed 
their evolution in these lowest metallicity galaxies).
\cite{nava06} revisited the observed N/O plateau with a large set of objects and determined a mean value for 
the N/O plateau of $-$1.43 with a standard deviation of $^{+0.071}_{-0.084}$.  
They further concluded from a $\chi^2$ analysis that only a small fraction of the observed scatter in N/O is intrinsic.

From the bottom panel of Figure~\ref{fig:NO}, we see that the sample assembled here also shows a plateau in N/O 
of log(N/O) = $-$1.56 $\pm$ 0.05.
The level of the plateau in our data is slightly lover than found by \citet{nava06}, but agrees fairly well with that found by \citet{izotov99}.
 While the observed dispersion is larger than that found for the blue compact dwarfs by \citet{izotov99},
 the \textit{intrinsic} dispersion agrees well for the two samples.
Clearly the relatively constant N/O value is a common characteristic of dwarf star forming galaxies, and not just
those undergoing a current burst of star formation.
\citet{vanzee06b} demonstrated that Leo~A, with 12 + log(O/H) = 7.38 $\pm$ 0.10 and log(N/O) = $-$1.53 $\pm$ 0.09, 
and GR~8, with 12 + log(O/H) = 7.65 $\pm$ 0.06 and log(N/O) = $-$1.51 $\pm$ 0.07, which are \textit{not} blue compact 
dwarf galaxies, are consistent with this plateau in log(N/O) at low values of O/H.  
However, both Leo~A and GR~8 have detailed star formation histories derived from \textit{Hubble Space Telescope} 
observations of their resolved stars which clearly show that the bulk of their star formation occurred well before 
the last 40 Myr \citep{tolstoy98, cole07, dohmpalmer98, weisz11}.
In fact, \citet{weisz11} show, from a nearly volume limited sample, that the majority of dwarf galaxies formed the 
bulk of their stellar mass prior to z $\sim$ 1, regardless of current morphological type. 
Since the low mass, metal-poor galaxies in the present sample and works cited appear to have nearly the same value of N/O, 
regardless of whether they have a current burst of star formation, it would seem that the young galaxy hypothesis is not 
a valid explanation for the plateau in N/O at low metallicity.

If the plateau in N/O is not due to young galaxy ages, what is its cause?  
Clearly nitrogen is behaving as a primary element at low metallicities. 
\citet{henry06} considered various scenarios and concluded that a wide range were consistent
with the observations.  At this point, a definitive explanation for the N/O plateau appears elusive.

%----------------------------------------------------------------------------------------------------------------------

\subsection{Best Estimate of Abundances}
\label{sec:best}

Determining an accurate and reliable oxygen abundance for an individual \ion{H}{2} 
region depends on measuring the combination of bright nebular and faint auroral 
emission lines (the ``direct" method).  
Many studies have emphasized that a ``direct'' abundance is not without 
systematic uncertainties.
Specifically, due to the high temperature sensitivity of the ``direct'' method,
inhomogeneous temperature distributions will lead to abundance underestimates.
The uncertainty in the absolute oxygen abundance determination by this method is 
$\sim$ 0.1 dex, but the error in relative metallicities is likely to be $<<$ 0.1 dex \citep{kewley08}.
However, \citet{bresolin2007} warns that T$_{e}$-based determinations only provide 
a lower limit if the temperature fluctuations are substantial.

In the absence of a temperature-sensitive auroral line detection, a mix of strong 
emission lines are used as a proxy for metallicity (strong-line methods: empirical, 
semi-empirical, and theoretical calibrations).
Strong-line calibrations are limited by sample selection effects, potentially making them appropriate for ranking objects
on a single scale, but not useful for determining an absolute metallicity as the 
various methods do not converge \citep[see e.g.,][]{yin07,kewley08,bresolin09,berg11}.  
If a strong-line method must be used, \citet{stasinska10} recommends only 
using a strong-line method for nebulae having {\it the same properties as those of the calibration sample}.

\citet[]{oey00, vanzee06a, vanzee06b, yin07, kewley08, perez-montero09, amorin10, 
moustakas10}, and others have investigated several strong-line calibrations including 
the O3N2 method, the N2 method, and the R$_{23}$ index, finding inconsistencies between 
methods that were largely related to variations in the hardness of
the ionizing radiation field, nitrogen abundance, and/or age of the stellar cluster.
There are several strong-line methods to chose from, but when compared they all
have similar uncertainties of 0.1-0.2 dex and discrepancies between them as large 
as 0.6 dex \citep[e.g.,][]{liang06,bresolin2007,yin07,kewley08}.
Improvements have been made in strong-line calibrations by the introduction of 
photoionization models to simultaneously fit the most prominent emission lines 
\citep[e.g.,][]{tremonti04,brinchmann04}.
However, \citet{yin07} found the MPA/JHU simultaneous line fitting SDSS abundances 
determined from the Charlot et al.\ (2006) photoionization models overestimate oxygen 
abundances by $\sim$0.34 dex compared to direct abundances.
They postulate the difference to be due to the models treatment of the onset of secondary 
nitrogen production, and thus could be eliminated with improved modeling.
One possible exception is the ONS calibration of \citet{pilyugin10}, for which they 
find deviations from T$_e$-based oxygen abundances of just $\sim$0.075 dex.

Here we investigate a subset of strong line abundances for our objects with ``direct"
abundances.
Following the methodology of \citet{berg11}, we calculated oxygen abundances from 
their strong lines for the 31 objects with ``direct" abundances listed in Table~\ref{tbl5}.
We determined abundances using the R$_{23}$ calibration of \citet{mcgaugh91}, 
the ONS calibration of \citet{pilyugin10},
and the N2 and O3N2 calibrations updated by \citet[][hereafter PMC09]{perez-montero09}.
The R$_{23}$ calibration of \citet{mcgaugh91} produces a bi-valued solution,
so to discriminate between the two branches \citet{mcgaugh94}, \citet{vanzee98},
and others advised using the ratio of I([\ion{N}{2}] $\lambda6584$)/I([\ion{O}{2}] $\lambda3727$).
\citet{mcgaugh94} suggested that [\ion{N}{2}]/[\ion{O}{2}] is approximately $< 0.1$ 
for low abundances and $> 0.1$ for high abundances, 
giving a rough distinction between lower and upper branches.
Using this distinction, we selected the appropriate branch calibration for each object.
Note that for metal poor objects with enhanced nitrogen, [\ion{N}{2}]/[\ion{O}{2}]
becomes a biased discriminator \citep[e.g.,][]{yin07,berg11,perez-montero2011}.
In a similar fashion, the ONS method of \citet{pilyugin10} requires two discriminators, 
[\ion{N}{2}] and [\ion{N}{2}]/[\ion{S}{2}], to distinguish between three classes of \ion{H}{2} regions.

We followed \citet{berg11} and assumed T$_{\rm e}=1.25\times10^4$ K to examine N/O 
ratios and calculate abundances with the N2 and O3N2 calibrations of PMC09.
This correction may be important for NGC 2537 and UGC 4393, which appear to
have somewhat discrepant nitrogen abundances (nitrogen enrichment for log(N/O) $>$ -1.0).
The other objects in this sample have average N/O ratios for their masses 
\citep[see e.g.,][]{berg11}.
The results are tabulated in Table~\ref{tbl8}. 

The mean offsets and dispersions relative to the direct abundances are
calculated and given at the bottom of Table~\ref{tbl8}.
Table~\ref{tbl8} shows that all four methods have significant dispersions,
with the ONS method showing the smallest dispersion (although larger than
anticipated) and the O3N2 method having the largest.  
The ONS method also has the smallest mean offset.  
Figure~7 presents a plot of differences between the R$_{23}$ and ONS method
abundances and the direct abundances as a function of abundance.  
This illustrates the results of Table~\ref{tbl8}, that the ONS method has a smaller
dispersion and a smaller mean offset from the direct method.
Thus, our data favor the ONS method, 
but do not support the claim of the very small error as found by \citet{pilyugin10}.
In Figure~7 we find no clear trend exists between the ``direct" method and the strong-line 
methods, implying that simple calibrations between methods are not possible.

With the relatively precise M-Z and L-Z relationships in place, and their
correspondingly low dispersions, oxygen abundances for normal (non-starburst) low luminosity 
galaxies can be inferred with relatively high confidence without a spectrum.
In fact, given reliable distance and photometry measurements, the resulting 
luminosity and mass estimates can be used as more reliable predictors of oxygen 
abundance than some strong-line calibrations.
As counter-intuitive as this idea may seem, it is a natural consequence of the inability of some 
strong-line methods to accurately predict the metallicity of individual \ion{H}{2} regions.
Studies of abundances in dwarfs which do not reproduce the L-Z and M-Z relationships, 
therefore, should raise suspicions concerning methodology.

%----------------------------------------------------------------------------------------------------------------------
%----------------------------------------------------------------------------------------------------------------------

\section{CONCLUSIONS}
\label{sec:conclusion}
We have determined uniform oxygen abundance metallicities for 31 
low luminosity galaxies in the \textit{Spitzer} LVL survey. 
With high-resolution spectral observations taken at the MMT, we were able to 
measure the intrinsically faint [\ion{O}{3}] $\lambda$4363 fluxes at strengths 
of 4$\sigma$ or greater and explicitly determine electron temperatures.
Metallicity measurements are important for characterizing many other properties, 
especially when the more reliable ``direct" method is used.
However, metallicity relationships tend to suffer from small number statistics in the low luminosity regime.
In particular, these measurements allowed us to better characterize the luminosity-metallicity and 
mass-metallicity relationships by doubling the number of reliable low-luminosity measurements.
We created a  ``Combined Select" sample of objects that have both reliable ``direct" oxygen abundance
determinations and distances estimated from the tip of the red giant branch or Cepheid variables.
With this sample, we find that both the luminosity-metallicity and the mass-metallicity relationships 
agree well with previous relationships defined for low luminosities.

From the 38 objects making up the ``Combined Select" sample, we found an optical L-Z relationship
of $12+\log(\mbox{O/H})=(6.27\pm0.21)+(-0.11\pm0.01)M_{B}$, with a dispersion of $\sigma=0.15$.
In comparison, the near-IR L-Z relationship for this data is 
$12+\log(\mbox{O/H})=(6.10\pm0.21)+(-0.10\pm0.01)M_{[4.5]}$, with a dispersion of $\sigma=0.14$.
While the slopes of the two L-Z relationships agree, our findings confirm the work of L06 in
that the near-IR relationship has lower scatter.
By converting NIR luminosity to a stellar mass estimate, we determined the M-Z relationship for our data
to be $12+\log(\mbox{O/H})=(5.61\pm0.24)+(0.29\pm0.03)M_{\star}$, with a dispersion of $\sigma=0.15$.
In agreement with the idea that mass is more fundamentally related to metallicity than luminosity, we find 
that the intrinsic scatter of the optical L-Z, NIR L-Z, and M-Z relationships decreases from 0.13 to 0.12 to 0.08.
However, the total dispersion of the M-Z relationship was measured to be no smaller than the L-Z relationships.
This suggests, given a reliable distance measurement and appropriate photometry,
luminosity is just as strong of a metallicity indicator as stellar mass.
Furthermore, with the dispersions in luminosity and mass roughly equal, either may be used in combination
with a reliable distance determination to estimate metallicity of a low luminosity dwarf with more 
confidence than when using strong-line calibrations.

Our observations of N/O abundances are in agreement with previous studies.
We find a positive correlation between N/O ratio and B-V color for 0.05 $\lesssim B-V \lesssim$ 0.75;
$\log(\mbox{N/O}) = (-1.92\pm0.08)+(1.18\pm0.19)\times(B-V)$, with a dispersion of $\sigma$ = 0.14.
Furthermore, in agreement with observations of blue compact galaxies, 
there are no objects with high N/O ratio (log(N/O) $>$ -1.4) below 12+log(O/H)=7.7.
Since the typical low luminosity galaxy in the Local Volume displays roughly constant star formation
over the age of the universe, the small dispersion in N/O at low values of O/H cannot be due to
the very recent birth of the galaxy.

%----------------------------------------------------------------------------------------------------------------------
%----------------------------------------------------------------------------------------------------------------------

\acknowledgements
DAB is grateful for support from a Penrose Fellowship and a NASA Space 
Grant Fellowship from the University of Minnesota.
EDS is grateful for partial support from the University of Minnesota.
We are grateful to the referee for a thorough analysis of this paper that 
greatly improved the analysis and presentation of this work.
Special thanks to John Moustakas and L. Andrew Helton for many
scientifically stimulating and helpful discussions.
Observations reported here were obtained at the MMT Observatory, a joint
facility of the Smithsonian Institution and the University of Arizona.
MMT observations were obtained as part of the University of Minnesota's 
guaranteed time on Steward Observatory facilities through membership in 
the Research Corporation and its support for the Large Binocular Telescope,
and granted by NOAO, through the Telescope System Instrumentation Program (TSIP).
TSIP is funded by the National Science Foundation.

This research has made use of NASA's Astrophysics Data System
Bibliographic Services and the NASA/IPAC Extragalactic Database
(NED), which is operated by the Jet Propulsion Laboratory, California
Institute of Technology, under contract with the National Aeronautics
and Space Administration.
This work was initiated as part of the Spitzer Space Telescope Legacy Science
Program and was supported by National Aeronautics and Space Administration 
(NASA) through contract 1336000 issued by the Jet Propulsion Laboratory (JPL), 
California Institute of Technology (Caltech) under NASA contract 1407.

%----------------------------------------------------------------------------------------------------------------------
%----------------------------------------------------------------------------------------------------------------------

\appendix

\section{Strong-Line Abundances for Galaxies Lacking Direct Abundances}
\label{sec:strong}

In Table~\ref{tbl9} we present strong-line abundances for the 12 objects 
in our sample without [\ion{O}{3}] $\lambda4363$ detections.
While these may not be as accurate as the direct abundances for the rest
of our sample, they may be useful for studies of these individual galaxies.
The O/H values derived using the ONS method for both these 12 objects (Table~\ref{tbl9})
and the objects with 	``direct abundances" (Table~\ref{tbl8}) are plotted in Figure~\ref{fig:strong}
where they are compared to our ``direct" abundances.
The two methods display coincident trends in metallicity with mass, yet the 
O/H abundances derived via the ONS calibration have a larger dispersion.
We have not conducted a statistical comparison, as not all galaxies have 
accurate distances, and the subset with accurate distance is quite small.

%----------------------------------------------------------------------------------------------------------------------

\section{70/160 $\mu$m Color Temperature-Metallicity Outliers}
\label{sec:outlier}

As noted in \S~\ref{sec:Sample}, two objects were of particular 
interest to this study (UGC 10818 and UGC 4393) because they appear
to be outliers from the global trend of 70/160 $\mu$m color temperature
as a function of metallicity as determined by \citet{engelbracht08}.
Specifically, based on \textit{Spitzer} observations of 66 starburst galaxies, they showed that the far-infrared color 
temperature of large dust grains increases toward lower metallicity down to 12 + log(O/H) $\sim$ 8.
However, the oxygen abundances found by \citet{engelbracht08} for these two objects were based on the
R$_{23}$ strong-line estimator. 
Our new spectroscopic results indicate that both UGC 4393 and UGC 10818 (SHOC 567) are near
the transition region between the upper and lower branches based on their [\ion{N}{2}]/[\ion{O}{2}] ratios,
and thus the R$_{23}$ method may not yield an accurate abundance for these systems.

While our observations of UGC 10818 are still ambiguous due to the degeneracy
in the strong-line metallicity calibrations, we derive an oxygen abundance
of 12 + log(O/H) = 7.82 based on the \citet{mcgaugh91} R$_{23}$ calibration.  
This increases the oxygen abundance of UGC 10818 by 0.51 dex compared to
previous measurements and moves UGC 10818 (SHOC 567) closer to the
original trend illustrated in \citet{engelbracht08}.  
Conversely, the ``direct" oxygen abundance of UGC 4393 was determined in this paper to be
12 + log(O/H) = 8.02 +/- 0.05, in agreement with the strong-line estimate presented in \citet{engelbracht08}.  
Thus, at first glance, these new observations appear to only impact the location of 
one of the two most extreme outliers in the original plot.

Perhaps more importantly, we have reproduced the 70/160 $\mu$m color temperature versus 12 + log(O/H) plot
of \citet{engelbracht08} with the addition of ``direct" abundance objects from this work in Figure~\ref{fig:temp}. 
Note that the star-bursting objects from \cite{engelbracht08} tend to have higher dust temperatures than
the low intensity objects studied in this paper. 
This may mean that the trend of increasing far-infrared dust temperature with decreasing metallicity was 
just a slice of a larger picture, where the selected samples were limited by star formation rates, 
which biased the view to a more narrow window.
With a more complete range of intensities in star forming galaxies now plotted, no clear trend emerges.

%----------------------------------------------------------------------------------------------------------------------
%----------------------------------------------------------------------------------------------------------------------

\newpage 

%----------------------------------------------------------------------------------------------------------------------
%----------------------------------------------------------------------------------------------------------------------

%Table 1 - Updated table values with data from Liese 06/25/11
\begin{deluxetable}{ccccccccccccc}
\tablewidth{0pt}
\rotate
\tabletypesize{\scriptsize}
\tablecaption{Low-Luminosity LVL Sample}
\setlength{\tabcolsep}{0.05in} 
\tablehead{
\CH{Galaxy} 	& \CH{RA}	& \CH{DEC} 	 & \CH{$F_{[4.5]}$} 	& \CH{$F_{K_S}$} 	&  \CH{D} 		& \CH{Reference} 	&  \CH{$M_{B}$} & \CH{$M_{[4.5]}$} & \CH{$M_{K_S}$} 	& \CH{$(B-V)_0$} 	& \CH{log $L_{[4.5]}$} 	& \CH{log $M_{\star}$}  \\
\CH{} 	     	& \CH{(J2000)} & \CH{(J2000)} & \CH{(mJy)} 	   	& \CH{(mJy)} 		& \CH{ (Mpc)}	& \CH{Method} 		& \CH{(mag)} 	   & \CH{(mag)} 		 & \CH{(mag)}   		& \CH{(mag)} 	& \CH{($L_\odot$)} 		& \CH{($M_\odot$)}}  
\startdata
UGC 521		& 00:51:12.1	&	12:01:26	& 2.24	& 5.43$\pm$2.88	& 10.9			& 13, v(flow)	& -15.16$\pm$0.50		& -17.93$\pm$0.61 		& -17.46$\pm$0.52		& 0.34$\pm$0.05		& 8.50$\pm$0.61 & 7.96$\pm$0.61	\\
UGC 695		& 01:07:46.4	&	01:03:52	& 3.02	& $< 8.45$		& 10.2			& 6, v(flow)	& -15.13$\pm$0.50		& -18.11$\pm$0.61		& -17.80$\pm$0.52		& 0.45$\pm$0.06		& 8.57$\pm$0.61 & 8.08$\pm$0.61	\\
NGC 404		& 01:09:27.0	&	35:43:05	& 239	& 676$\pm$340	& 3.05$\pm$0.04	& 4, trgb		& -16.39$\pm$0.07		& -20.23$\pm$0.35		& -19.94$\pm$0.14		& 0.83\TN{4}			& 9.42$\pm$0.35 & 9.20$\pm$0.35	\\
UGC 1056	& 01:28:47.6	&	16:41:21	& 4.14	& 29.1$\pm$4.0	& 10.32			&  6, v(flow)	& -15.09$\pm$0.52		& -18.47$\pm$0.61		& -19.17$\pm$0.50		& 0.57$\pm$0.07		& 8.72$\pm$0.61 & 8.62$\pm$0.61	\\
UGC 1176	& 01:40:09.9	&	15:54:20	& 6.22	& $< 26.6$		& 9.04$\pm$1.66	& 9, bs		& -15.48$\pm$0.93		& -18.63$\pm$0.98		& -18.78$\pm$0.92		& 0.31$\pm$0.10		& 8.78$\pm$0.98 & 8.48$\pm$0.98	\\
NGC 784		& 02:01:16.9	&	28:50:09	& 35.0	& 72.2$\pm$9.9	& 5.19$\pm$0.12	& 12, trgb		& -16.50$\pm$0.12		& -19.30$\pm$0.36		& -18.66$\pm$0.20		& 0.40$\pm$0.04		& 9.05$\pm$0.36 & 8.48$\pm$0.36	\\
UGC 2716	& 03:24:08.1	&	17:45:15	& 7.58	& 22.2$\pm$4.7	& 6.2				& 6, v(flow)	& -15.31$\pm$0.50		& -18.04$\pm$0.61		& -17.78$\pm$0.51		& 0.31$\pm$0.06		& 8.55$\pm$0.61 & 8.13$\pm$0.61	\\
KKH 37		& 06:47:45.4	&	80:07:26	& 1.58	& 6.43$\pm$2.81	& 3.39$\pm$0.12	& 4, trgb		& -11.98$\pm$0.20		& -15.01$\pm$0.39		& -15.11$\pm$0.20		& 0.54$\pm$0.06		& 7.34$\pm$0.39 & 7.01$\pm$0.39   \\
\enddata
\end{deluxetable}

%----------------------------------------------------------------------------------------------------------------------

% Table 2
\begin{deluxetable}{cccccc}
\tabletypesize{\footnotesize}
\tablewidth{0pt}
\tablecaption{Observing Log\label{tab:slits}}
\tablehead{
\CH{Slit } 		& \CH{RA} 	&\CH{Dec}	& \CH{PA} 	& \CH{Run}	& \CH{T$_{\rm int}$} \\
 \CH{Position}	& \CH{(2000)} 	& \CH{(2000)} 	& \CH{(deg)}	& \CH{}		& \CH{(sec)}
}
\startdata
UGC  521-A      		& 00:51:11.9  	&	12:01:34 	&  -55.71 	& Nov08 	& 3 $\times$ 900 \\
UGC  521-B		& 00:51:12.1	&   	12:01:31 	&  -55.71 	& Nov08 	& 2 $\times$ 900 \\
UGC  695-E 		& 01:07:46.5 	& 	01:03:53 	&   29.44 	& Jan10 	& 4 $\times$ 900 \\
NGC  0404-A		& 01:09:26.0	&       35:43:00	& -76.70	& Jan10	& 3 $\times$ 600 \\
UGC  1056-A 		& 01:28:47.3	&  	16:41:16	&   45.19 	& Jan10 	& 3 $\times$ 900 \\
UGC  1056-B 		& 01:28:47.5	&  	16:41:21 	&   45.19 	& Jan10 	& 4 $\times$ 900 \\
UGC  1176-A 		& 01:40:11.9 	&	15:54:46 	&   42.14 	& Jan10 	& 4 $\times$ 900 \\
UGC  784-B 		& 02:01:16.5	&  	28:50:06 	&   52.76 	& Jan10 	& 4 $\times$ 900 \\
\enddata
\tablecomments{The low-luminosity LVL sample observing log.
Galaxy name and \ion{H}{2} region label are listed in Column 1.
The right ascension and declination of the individual \ion{H}{2} regions are given in units of
hours, minutes, and seconds, and decrees, arcminutes, and arcseconds respectively.
The position angle (PA) gives the rotation of the slit counter clockwise from North.
Note that the RA and Dec positions for ``A" regions are accurate and denote the center slit placement.
However, ``B" and ``C" regions only have estimated RA and Dec positions, as these are just extra 
extractions along the same slit that is aligned through ``A".}
\label{tbl2}
\end{deluxetable}

%----------------------------------------------------------------------------------------------------------------------

% Table 3, new:
\begin{deluxetable}{lcccccccccc}
\tabletypesize{\scriptsize}
\rotate
\setlength{\tabcolsep}{0.04in} 
\tablecaption{ Emission-Line Intensities and Equivalent Widths for Low-Luminosity LVL Galaxies}
\tablewidth{0pt}
\tablehead{
\CH{} & \multicolumn{10}{c}{$I(\lambda)/I(\mbox{H}\beta)$} }
\startdata
{Ion} 						& {UGC}   			& {UGC}   			& {UGC}    		& {UGC}    		& {UGC}			& {NGC} 			& {NGC}			& {UGC}    		& {KKH}   			& {NGC}    			\\
						& {521 A}			& {695 E} 			& {1056 A}		& {1056 B} 		& {1176 A}		& {784 A} 			& {784 B}			& {2716 A} 		& {037 A} 			& {2537 A} 			\\
\hline																																													
{[O II] $\lambda$3727}		& 1.59$\pm$0.03   	& 3.25$\pm$0.07   	& 3.32$\pm$0.13	& 2.80$\pm$0.06     	& 2.27$\pm$0.05   	& 2.19$\pm$0.06	& 2.61$\pm$0.07   	& 2.08$\pm$0.04   	& 6.51$\pm$0.08   	& 3.42$\pm$0.07   		\\
{He I $\lambda$3820}		& \nodata        		& \nodata         		& \nodata         		& \nodata         		& \nodata         		& \nodata   		& \nodata         		& \nodata         		& \nodata         		& \nodata         			\\
{H9 $\lambda$3835}		& 0.07$\pm$0.02	& 0.11$\pm$0.01   	& \nodata         		& \nodata         		& 0.07$\pm$0.01   	& 0.04$\pm$0.01  	& 0.07$\pm$0.01   	& 0.08$\pm$0.02   	& \nodata         		& \nodata         			\\
{[Ne III] $\lambda$3868}		& 0.33$\pm$0.02	& 0.19$\pm$0.01   	& 0.27$\pm$0.04   	& 0.30$\pm$0.02   	& 0.28$\pm$0.01   	& 0.33$\pm$0.01   	& 0.29$\pm$0.01   	& 0.38$\pm$0.02   	& \nodata         		& 0.15$\pm$0.01   		\\
{He I+H8 $\lambda$3889}	& 0.17$\pm$0.02	& 0.25$\pm$0.01   	& 0.17$\pm$0.03   	& 0.24$\pm$0.02   	& 0.19$\pm$0.01   	& 0.18$\pm$0.01	& 0.23$\pm$ 0.01  	& 0.26$\pm$0.01   	& \nodata         		& 0.17$\pm$0.01   		\\
{[Ne III]+H7 $\lambda$3968}	& 0.34$\pm$0.02	& 0.36$\pm$0.01   	& 0.37$\pm$ 0.02  	& 0.54$\pm$0.02   	& 0.59$\pm$0.01   	& 0.20$\pm$0.01	& 0.51$\pm$ 0.01  	& 0.32$\pm$0.01   	& \nodata         		& 0.30$\pm$0.01   	  	\\
\enddata
\tablecomments{Optical line fluxes for \ion{H}{2} regions measured from the MMT spectra using deblended Gaussian fits 
and multiple component fits when necessary. Fluxes are relative to H$\beta$ = 1.00 and are corrected for reddening. 
The H$\beta$ flux is given for reference, with units of $10^{-17}$ erg s$^{-1}$ cm$^{-2}$. EWs are given in units of \AA.
Multiple extractions are listed for those objects which had more than one bright H$\alpha$ region aligned in the slit;
letter attached to the object name specifies the H$\alpha$ region label for that extraction (see Figure~\ref{fig1}).
Note that uncertainties listed in this table reflect the statistical uncertainties in the flux through the slit only, and do not account for slit loses.} 
\label{tbl3}
\end{deluxetable}

%----------------------------------------------------------------------------------------------------------------------

% Table 4: 
% Only objects with an [O III] 4363 strength greater than 4 sigma are listed here:
\begin{deluxetable}{cccccccccccc}
\tabletypesize{\scriptsize}
\rotate
\tablenum{4}
\tablecaption{Ionic and Total Abundances \label{tbl4}}
\tablewidth{0pt}
\tablehead{
\CH{Galaxy} & \CH{H$\alpha$ Region} & \CH{t$_2$} & \CH{t$_3$} & \CH{O$^+$/H$^+$} & \CH{O$^{++}$/H$^+$} & \CH{O/H} & \CH{12 + log(O/H)} & \CH{N$^{+}$/H$^{+}$} & \CH{log(N/O)} & \CH{N/H}  \\
\CH{}       	    & \CH{}   & \CH{(K)}      & \CH{(K)}      & \CH{($\times10^5$)}  &  \CH{($\times10^5$)}     &  \CH{($\times10^5$)}  & \CH{(dex)} &  \CH{($\times10^6$)} & \CH{(dex)} & \CH{($\times10^6$)}}  
\startdata  		
UGC 521           & A   & 14200$\pm$900           	& 16500$\pm$1100           & 1.59$\pm$0.32 & 3.05$\pm$0.42 &        4.64$\pm$0.53 &      7.67$\pm$0.05 & 0.29$\pm$0.06      & -1.61$\pm$0.07     & 1.15$\pm$0.25      \\
UGC 695		& E	& 14000$\pm$1900		& 15800$\pm$2200		& 3.45$\pm$1.48 & 1.50$\pm$0.45 &	4.95$\pm$1.55	&	7.69$\pm$0.12	& 1.07$\pm$0.29	& -1.49$\pm$0.04	& 1.61$\pm$0.53	\\
UGC 1056  	& A	& 13000$\pm$2500		& 13500$\pm$2600		& 4.50$\pm$2.89 & 3.26$\pm$1.56 &	7.75$\pm$3.29	&	7.89$\pm$0.15	& 1.48$\pm$0.59	& -1.48$\pm$0.05	& 2.57$\pm$1.14	\\
UGC 1056  	& B	& 12600$\pm$900 		& 12700$\pm$900		& 4.23$\pm$1.06 & 5.43$\pm$1.04 &	9.66$\pm$1.49	&	7.98$\pm$0.06	& 1.34$\pm$0.21	& -1.49$\pm$0.03	& 3.16$\pm$0.52	\\
UGC 1176	& A	& 12600$\pm$500 		& 12800$\pm$500		& 3.40$\pm$0.47 & 5.91$\pm$0.59 &	9.31$\pm$0.75	&	7.97$\pm$0.03	& 1.29$\pm$0.11	& -1.40$\pm$0.02	& 3.73$\pm$0.34	\\
NGC 784		& B	& 12900$\pm$600		& 13400$\pm$700		& 3.57$\pm$0.59 & 4.70$\pm$0.58 &	8.27$\pm$0.83	&	7.92$\pm$0.04	& 0.78$\pm$0.08	& -1.63$\pm$0.02	& 1.94$\pm$0.22	\\
NGC 784		& A	& 12500$\pm$500		& 12400$\pm$500		& 3.41$\pm$0.48 & 7.20$\pm$0.78 &      10.61$\pm$0.91	&	8.03$\pm$0.04	& 1.24$\pm$0.11	& -1.44$\pm$0.02	& 3.87$\pm$0.36	\\
UGC 2716	& A	& 12800$\pm$500 		& 13100$\pm$500		& 2.96$\pm$0.42 & 6.41$\pm$0.70 &	9.37$\pm$0.82	&	7.97$\pm$0.04	& 0.99$\pm$0.09	& -1.47$\pm$0.02	& 3.14$\pm$0.31	\\
\enddata
\tablecomments{Electron temperatures and ionic and total abundances for objects with an [\ion{O}{3}] $\lambda4363$ line signal to noise ratio of $4\sigma$ or greater.
Electron temperatures were calculated using the  [\ion{O}{3}] ($\lambda4959 + \lambda5007$)/$\lambda4363$ diagnostic line ratio.} 
\end{deluxetable}

%----------------------------------------------------------------------------------------------------------------------

% Table 5: 
\begin{deluxetable}{ccccccc}
\tabletypesize{\scriptsize}
\tablenum{5}
\tablecaption{Error Weighted Average Abundances \label{tbl5}}
\tablewidth{0pt}
\tablehead{
\CH{Galaxy} 	& \CH{``Select"} 	&	\CH{12 + log(O/H)} & \CH{[O~III]/[O~II]} 	& \CH{log(N/O)} 	& \CH{Previous Literature} \\
\CH{}            	& \CH{Sample?}     	&	\CH{(dex)}                & \CH{}			     		& \CH{(dex)} 		& \CH{Abundances?}}
\startdata 
UGC 521           & 			& 7.67$\pm$0.05 	& 2.29$\pm$0.06	& -1.61$\pm$0.07 	& D: 12, 13     		\\
UGC 695		&			& 7.69$\pm$0.12	& 0.50$\pm$0.01	& -1.49$\pm$0.04 	& 				\\
UGC 1056	&			& 7.97$\pm$0.06	& 0.91$\pm$0.02	& -1.49$\pm$0.02 	&				\\
UGC 1176	&			& 7.97$\pm$0.05	& 1.59$\pm$0.05	& -1.40$\pm$0.02 	&				\\
NGC 784		& $\surd$		& 7.97$\pm$0.06	& 1.47$\pm$0.04	& -1.54$\pm$0.10 	& S: 2,6			\\
UGC 2716	&			& 7.97$\pm$0.05	& 2.05$\pm$0.06	& -1.47$\pm$0.02 	&				\\
NGC 2537	&			& 8.14$\pm$0.13	& 0.63$\pm$0.02	& -1.04$\pm$0.04 	& S: 2, 21-23		\\
UGC 4278	&			& 7.69$\pm$0.05	& 1.08$\pm$0.02	& -1.60$\pm$0.03 	& D: 16, 17 		\\
\enddata
\tablecomments{
For the 10 objects with multiple \ion{H}{2} regions containing strong [\ion{O}{3}] $\lambda$4363, 
error weighted averages were used to determine best estimates of relative abundances and oxygen abundances.
Column (2) highlights our ``Select" sample. 
Columns (3) - (5) lists the the new ``direct" oxygen abundances, ionization strengths, and nitrogen abundances relative to oxygen determined by this work.
Average values were determined using a weight of $1/\sigma_i^2$ for each component, where uncertainties represent the 
standard deviation of the weighted mean or the weighted dispersion, which ever is greater.
Column (6) shows which objects have previous oxygen abundance determinations in the literature, where `D' is noted for objects with ``direct" oxygen 
abundances, while `S' indicates objects with strong-line abundances.
Note that we are providing ``direct" oxygen abundances for the first time for 19 of these objects.} 
\tablerefs{
(1) \citet{kniazev03};
(2) \citet{moustakas06};
(3) \citet{gallagher89};
(4) \citet{vaduvescu07};
(5) \citet{miller96};
(6) \citet{hunter85};
(7) \citet{pustilnik05};
(8) \citet{croxall09};
(9) \citet{kewley05};
(10) \citet{kinman81}; 
(11) \citet{hopp91};
(12) \citet{vanzee97b};
(13) \citet{vanzee97a};
(14) \citet{izotov07b};
(15) \citet{liang07};
(16) \citet{kniazev04};
(17) \citet{izotov06};
(18) \citet{hunter99};
(19) \citet{hunter82};
(20) \citet{martin97};
(21) \citet{engelbracht08}; 
(22) \citet{gildepaz00b}; 
(23) \citet{gildepaz00a};
(24) \citet{mccall85};
(25) \citet{kobulnicky99};
(26) \citet{sabbadin84};
(27) \citet{matteucci85}
}
\end{deluxetable}

%----------------------------------------------------------------------------------------------------------------------

% Table 6: 
\begin{deluxetable}{ccccccccccccc}
\tablewidth{0pt}
\rotate
\tabletypesize{\scriptsize}
\tablenum{6}
\tablecaption{Additional ``Select" Galaxies from the Literature \label{tbl6}}
\tablehead{
\CH{Galaxy} 	& \CH{RA} 	& \CH{DEC} 	& \CH{$F_{[4.5]}$} 	& \CH{$F_{K_S}$} 	&  \CH{D} 	 	& {Ref}	&  \CH{$M_{B}$} 	& \CH{$M_{[4.5]}$} 	& \CH{$M_{K_S}$} 	& \CH{$(B-V)_0$} 		& \CH{log $L_{[4.5]}$} 	& \CH{log $M_{\star}$}  \\
\CH{} 		& \CH{(J2000)} & \CH{(J2000)}	& \CH{(mJy)} 		& \CH{mJy} 		& \CH{ (Mpc)} 	& {}		& \CH{(mag)}		& \CH{(mag)} 		& \CH{(mag)} 		& \CH{mag} 		& \CH{($L_\odot$)} 		& \CH{($M_\odot$)}}  
\startdata
\multicolumn{13}{c}{L06 ``Select" Objects} \\
\hline \\
WLM			& 00:01:58.6	& -15:27:12	& 62.9			& 117			& 0.97$\pm$0.02	& 1	& -13.50$\pm$0.05	& -16.29$\pm$0.34	& -15.54$\pm$0.34	& 0.46$\pm$0.03	& 7.85$\pm$0.34		& 7.19$\pm$0.34	\\ %1
NGC 55		& 00:14:53.6	& -39:11:48	& 1390			& 2630			& 2.11$\pm$0.04	& 2	& -18.20$\pm$0.11	& -21.34$\pm$0.36	& -20.61$\pm$0.16	& 0.55$\pm$0.08\TN{A} & 9.87$\pm$0.35		& 9.30$\pm$0.35	\\ %2
UGC 00668	& 01:04:49.1	& 02:07:31	& 90.1			& 232			& 0.75$\pm$0.02	& 1	& -13.61$\pm$0.14	& -16.13$\pm$0.37	& -15.73$\pm$0.27	& 0.40$\pm$0.04	& 7.78$\pm$0.37		& 7.14$\pm$0.37	\\ %3
NGC 1705	& 04:54:13.7	& -53:21:41	& 19.3			& 44.4			& 5.11$\pm$0.17	& 3 	& -15.77$\pm$0.52	& -18.62$\pm$0.60	& -18.10$\pm$0.57	& 0.38$\pm$0.18\TN{A} & 8.78$\pm$0.60		& 8.19$\pm$0.60	\\ %4
NGC 2366	& 07:28:49.6	& 69:12:32	& 4.99			& 110			& 3.21$\pm$0.05	& 2	& -15.95$\pm$0.11	& -18.64$\pm$0.35	& -18.08$\pm$0.33	& 0.32$\pm$0.05	& 8.79$\pm$0.35		& 8.15$\pm$0.35	\\ %5
UGC 4305	& 08:19:09.0	& 70:43:28	& 64.6			& 216			& 3.38$\pm$0.05	& 2	& -16.11$\pm$0.12	& -19.03$\pm$0.31	& -18.92$\pm$0.23	& 0.19$\pm$0.06	& 8.95$\pm$0.31		& 8.48$\pm$0.31	\\
\enddata
\tablecomments{On top are galaxies taken from L06 that meet the ``Select" specifications listed in Section~\ref{sec:Gold}. The bottom half lists the additional galaxies taken from \citet{marble10} and \citet{vanzee06a} that meet the ``Select" specifications of having oxygen abundances based on [O~III] $\lambda4363$ measurements of strength 4 $\sigma$ or greater and secure distance measurements. All distances were determined using the tip of the red giant branch, with the exception of Leo A, which used Cepheid variables. $(B-V)_0$ optical colors are from Liese \cite{vanzee12}, unless otherwise noted.}
\tablerefs{
(1) \citet{rizzi07};
(2) \citet{dalcanton09};
(3) \citet{tosi01};
(4) \citet{dolphin03};
(5) \citet{tully06};
(6) \citet{sanna08};
(7) \citet{cannon03};
(8) \citet{sakia04};
(9) \citet{karachentsev03};
}
\end{deluxetable}

%----------------------------------------------------------------------------------------------------------------------

% Table 7: 
\begin{deluxetable}{cccc}
\tabletypesize{\scriptsize}
\tablenum{7}
\tablecaption{Abundances for Additional ``Select" Galaxies \label{tbl7}}
\tablewidth{0pt}
\tablehead{
\CH{Galaxy} 	&	\CH{12 + log(O/H)} 	& \CH{log(N/O)} 	& \CH{Reference} \\
\CH{}            	&	\CH{(dex)}             	& \CH{(dex)}  		& \CH{} }
\startdata 
WLM			& 7.83$\pm$0.06	& -1.49$\pm$0.01	& 1	\\ %1
NGC 55		& 8.05$\pm$0.10	& -1.26$\pm$0.05	& 2 \\ %2
UGC 00668	& 7.62$\pm$0.05	& -1.51$\pm$0.10	& 3 \\ %3
NGC 1705	& 8.21$\pm$0.05	& -1.75$\pm$0.06	& 4 \\ %4
NGC 2366	& 7.91$\pm$0.05	& -1.17$\pm$0.26	& 5 \\ %5
UGC 4305	& 7.92$\pm$0.10	& -1.52$\pm$0.11	& 6 \\ %6
UGC 4459	& 7.82$\pm$0.09	& -1.32$\pm$0.17	& 7 \\ %7
Leo A		& 7.30$\pm$0.05	&  -1.53$\pm$0.09	& 8 \\
\enddata
\tablecomments{The top portion of the table lists the objects  and their abundances which were included in the L06 sample.
The bottom half lists additional objects found in the literature. All objects meet the ``Select" sample criteria.} 
\tablerefs{
(1) \citet{lee05};
(2) \citet{tullman03};
(3) \citet{lee03a};
(4) \citet{lee04};
(5) \citet{saviane08};
(6) \citet{lee03b};
(7) \citet{croxall09};
(8) \citet{vanzee06b};
(9) \citet{vanzee00};
(10) \citet{kniazev05};
(11) \citet{ks96};
(12) \citet{russell90};
(13) \citet{vanzee06a};
(14) \citet{skillman03}
(15) \citet{thuan05};
(16) \citet{vanzee97a};
}
\end{deluxetable}

%----------------------------------------------------------------------------------------------------------------------

% Table 8
\begin{deluxetable}{cccccc}
\tabletypesize{\scriptsize}
\tablenum{8}
\tablecaption{Strong-Line Oxygen Abundances for Our ``Direct" Detection Galaxies \label{tbl8}}
\tablewidth{0pt}
\tablehead{
\CH{}       		& \CH{}		& \multicolumn{4}{c}{12 + log(O/H)} \\
\cline{3-6} \\
\CH{Galaxy} 	& \CH{[\ion{N}{2}]/[\ion{O}{2}]} 	& \CH{R$_{23}$}          & \CH{ONS}	& \CH{N2}       & \CH{O3N2} }  
\startdata 
UGC 521 A		&	0.03	&	7.90	(L)	& 7.65 (3)	&	7.87	&	8.14	        	\\
UGC 695	E		&	0.04	&	8.01	(L)	& 7.93 (3)	&	8.13	&	8.34		\\
UGC 1056 A		&	0.04	&	8.08	(L)	& 8.03 (3)	&	8.15	&	8.29		\\
UGC 1176 A		&	0.05	& 	8.03 (L)	& 7.90 (3)	&	8.04	&	8.18		\\
NGC 784 A		&	0.05	&	8.08 (L)	& 7.95 (3)	&	8.02	&	8.15		\\
UGC 2716 A		&	0.04	&	8.08	(L)	& 7.93 (3)	& 	7.99	&	8.15		\\
NGC 2537 A		&	0.14	&	8.58	(U)	& 8.41 (2)	&	8.27	&	8.29		\\
UGC 4278 A		&	0.03	& 	7.78 (L)	& 7.64 (3)	&	7.88	&	8.19		\\
\enddata
\tablecomments{
Strong-line abundances are calculated for the 31 objects with spectra containing an [O~III] $\lambda4363$ signal-to-noise of 4 or greater.
Four strong-line calibrations are given: 1) The R$_{23}$ method of \citet{mcgaugh91} is listed in column (3), 
where the appropriate branch was selected based on the [N~II]/[O~II] ratio given in column (2), with [N~II]/[O~II]$\approx0.1$ drawing the division.
Branch selection is denoted by (L) for lower branch and (U) for upper branch.
2) The ONS calibration of \citet{pilyugin10} is given in column (4).
\citet{pilyugin10} divides their calibration into 3 classes of \ion{H}{2} regions, where we have used
(1) for regions with $\log{(N_2)} > -0.1$,
(2) for $\log{(N_2)} < -0.1$ and $\log{(N_2/S_2)} > -0.25$,
and (3) for $\log{(N_2)} < -0.1$ and $\log{(N_2/S_2)} < -0.25$.
3) The N2 calibration, corrected for N/O ratio variations by PMC09, is given in column (5).
4) The O3N2 calibration, also corrected for N/O ratio variations by PMC09, is given in column (6).
The last two rows give the average offset and dispersion of the strong-line abundances from their ``Direct" abundance counterparts.}  
\end{deluxetable}

% Table 9
\begin{deluxetable}{ccccccc}
\tabletypesize{\scriptsize}
\tablenum{9}
\tablecaption{Strong-Line Oxygen Abundances for [\ion{O}{3}] $\lambda4363$ Non-Detection Objects \label{tbl9}}
\tablewidth{0pt}
\tablehead{
\CH{}       		& \CH{}			       		& \multicolumn{4}{c}{12 + log(O/H)} \\
\cline{3-6} \\
\CH{Galaxy} 	& \CH{[\ion{N}{2}]/[\ion{O}{2}]} 	& \CH{$R_{23}$}  & \CH{ONS}	& \CH{N2}       & \CH{O3N2} }  
\startdata 
NGC 404			& 	0.04	&	7.23 (L)	& 7.55 (3)	& 	8.53	&	8.56 		\\
KKH 037			& 	0.06	&	8.32 (U)	& 8.33 (1)	& 	8.53	&	8.56	 	\\
CGCG 035-007 	& 	0.05	&	8.00 	(L)	& 8.10 (3)	&  	8.57 	&	8.37	        	\\	
UGC 5672		&	0.06	& 	8.00	(L)	& 8.26 (3)	&	8.37	&	8.29		\\
UGC 5692 		&	0.15	&	8.40 	(U)	& 8.08 (3)	&	8.53	&	8.31		\\
UGC 6782 		&	0.08	& 	7.85	(L)	& \nodata	&	8.18	&	8.27		\\
UGC 6900 		&	0.12	& 	8.00 	(U)	& 7.69 (3)	&	8.08	&	8.17		\\
UGC 7599		&	\nodata &	8.09 (U)	& \nodata	& \nodata	& \nodata		\\
\enddata
\tablecomments{Strong-line abundances are calculated for HII regions with no [\ion{O}{3}] $\lambda4363$ detections.
Most objects in our sample have at least one HII region with a ``direct" abundance determination, 
but 12 objects have good spectra and no [O~III] $\lambda4363$ measurement. 
Four strong-line calibrations are given for these objects: 1) The R$_{23}$ method of \citet{mcgaugh91} is listed in column (3), 
where the appropriate branch was selected based on the [N~II]/[O~II] ratio given in column (2). [N~II]/[O~II]$\approx0.1$ draws the branch division.
Branch selection is denoted by (L) for lower branch and (U) for upper branch.
2) The ONS calibration of \citet{pilyugin10} is given in column (4).
\citet{pilyugin10} divides their calibration into 3 classes of HII regions, where we have used
(1) for regions with $\log{(N_2)} > -0.1$,
(2) for $\log{(N_2)} < -0.1$ and $\log{(N_2/S_2)} < -0.25$,
and (3) for $\log{(N_2/S_2)} > -0.25$.
3) The N2 calibration, corrected for N/O ratio variations by PMC09, is given in column (5).
4) The O3N2 calibration, also corrected for N/O ratio variations by PMC09, is given in column (6).}  
\end{deluxetable}

%----------------------------------------------------------------------------------------------------------------------

% Figure 1A:
\begin{figure}
\epsscale{1.0}
\plotone{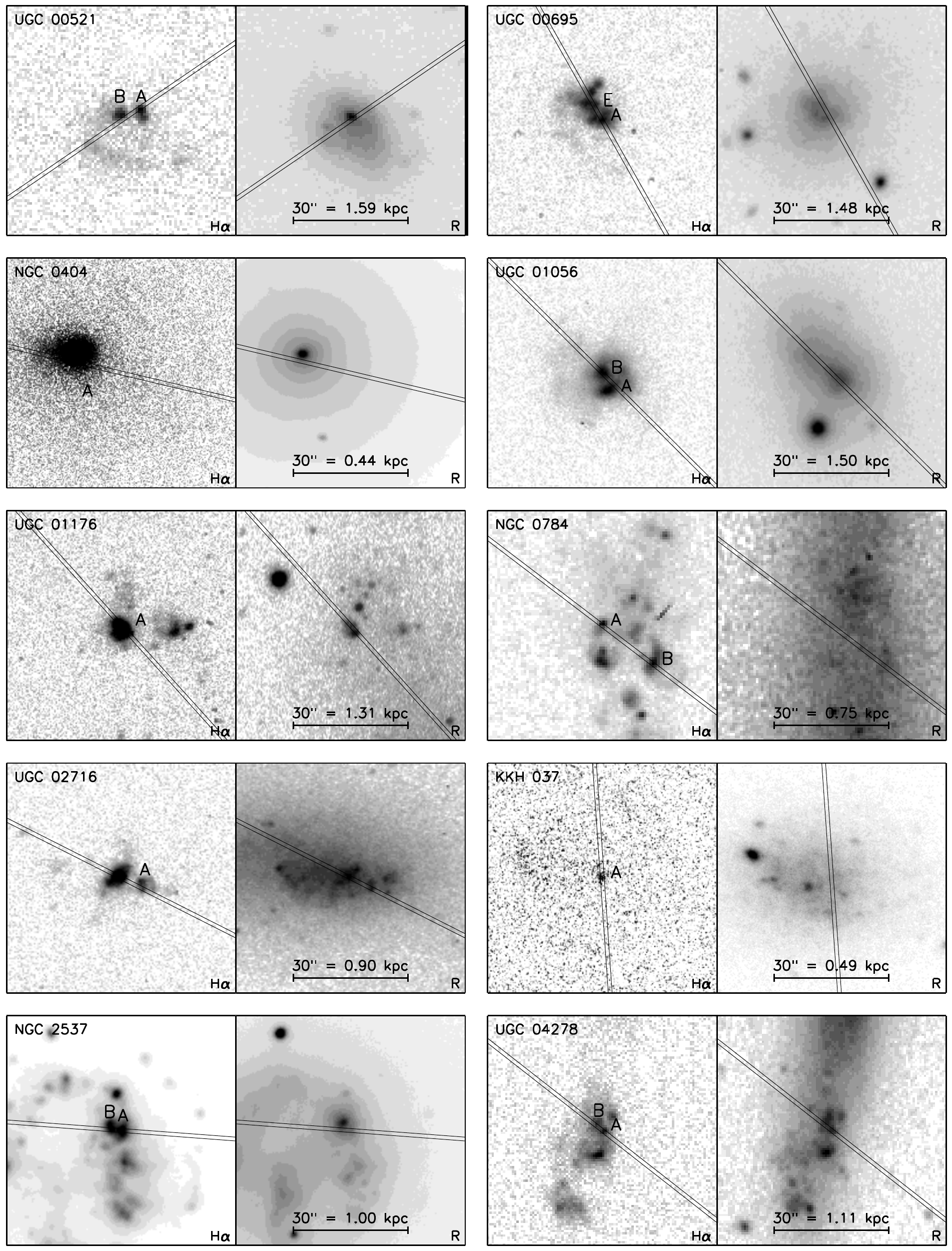}
\label{fig:objects}
\end{figure}

% Figure 1B:
\begin{figure}
\epsscale{1.0}
\plotone{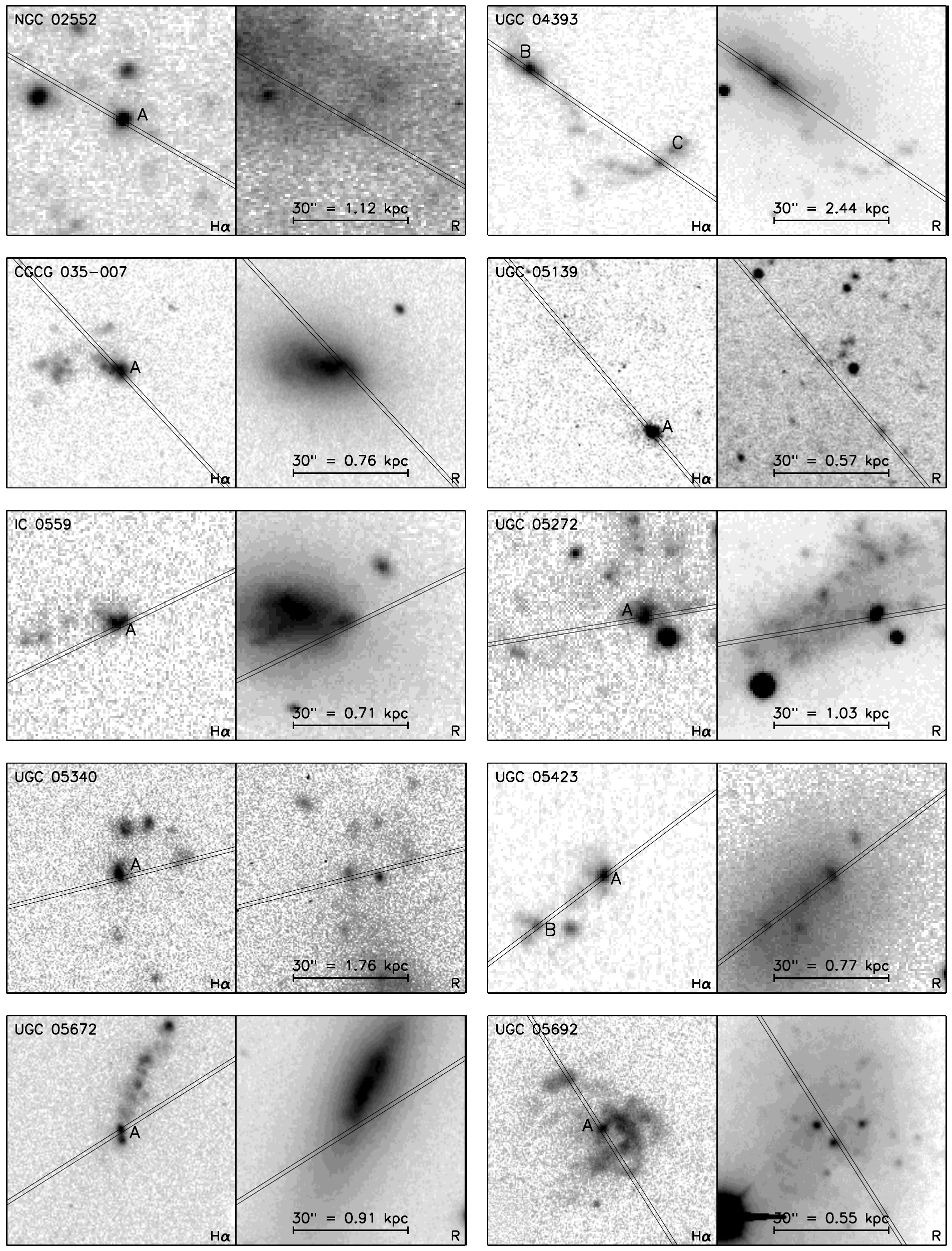}
\end{figure}

% Figure 1C
\begin{figure}
\epsscale{1.0}
\plotone{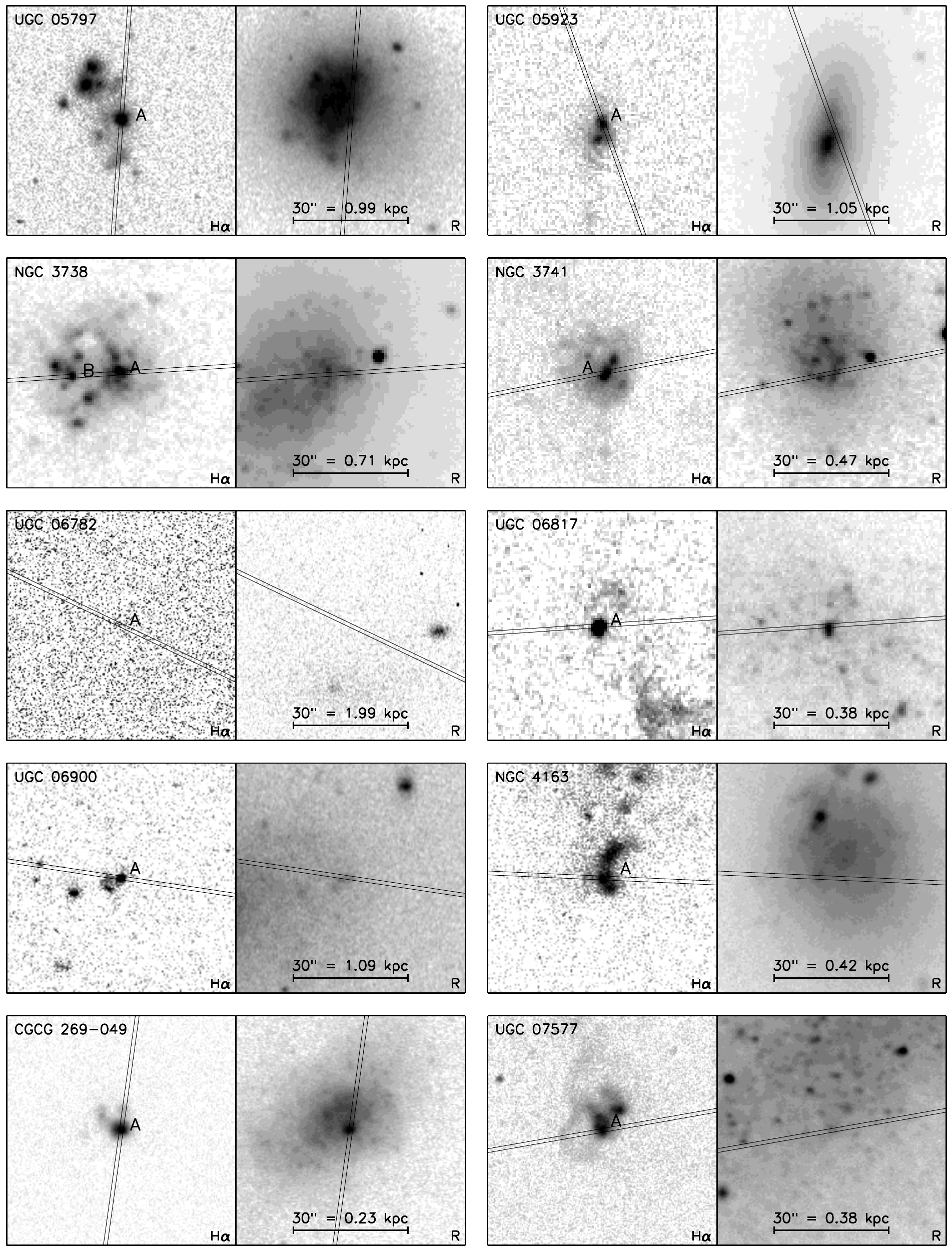}
\end{figure}

% Figure 1D:
\begin{figure}
\epsscale{1.0}
\plotone{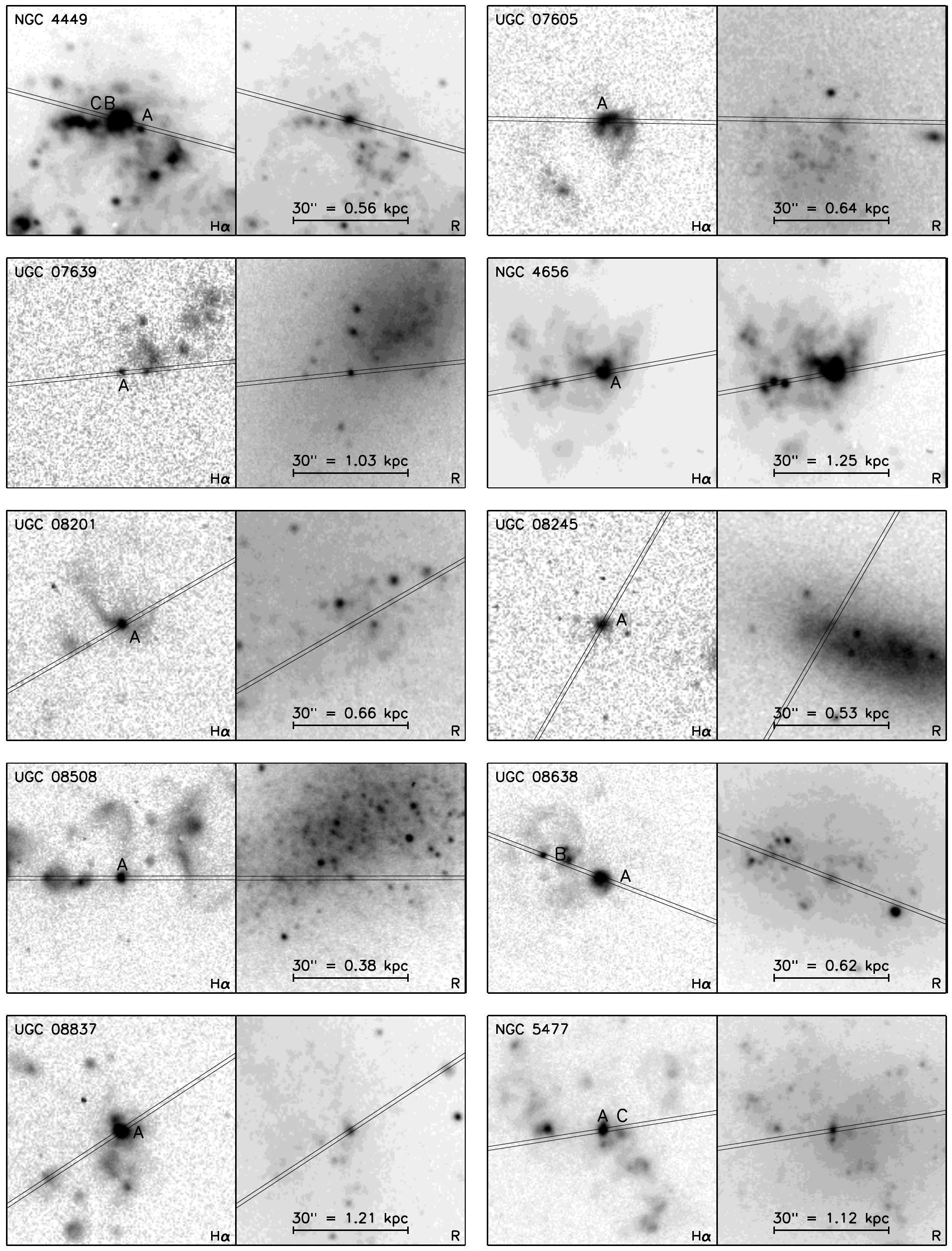}
\end{figure}

% Figure 1E:
\begin{figure}
\epsscale{1.0}
\plotone{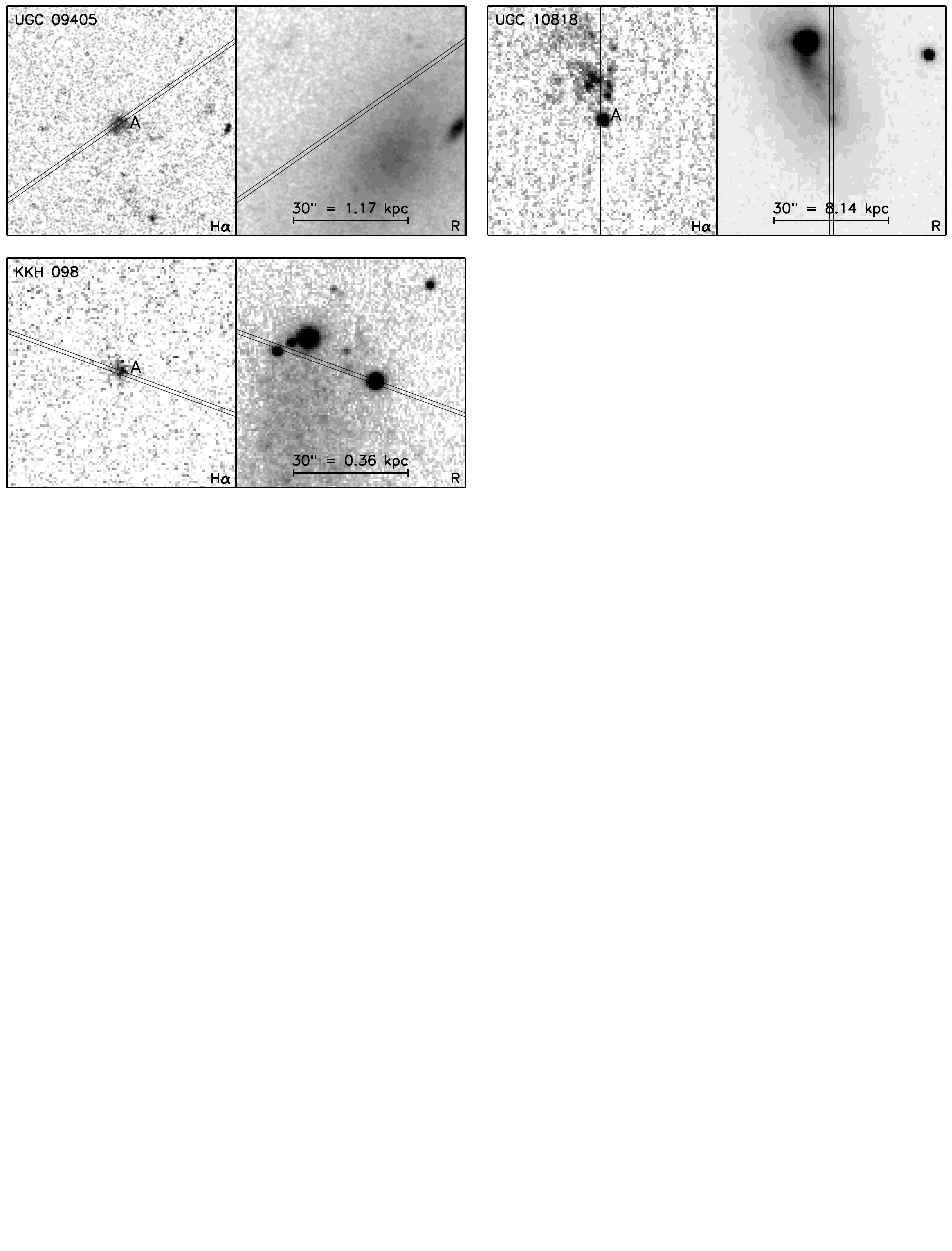}
\caption{H$\alpha$ and R-band images of the objects in the present low-luminosity LVL sample.
The angular scale of the images is 60\arcsec $\times$ 60\arcsec with North directly up and East to the left.
The line across the images represents the slit position during observation. The brightest \ion{H}{2}
regions are labeled with letters. See Table 1 for more details.}
\label{fig1}
\end{figure}

%----------------------------------------------------------------------------------------------------------------------

\begin{figure}
\epsscale{0.9}
\plotone{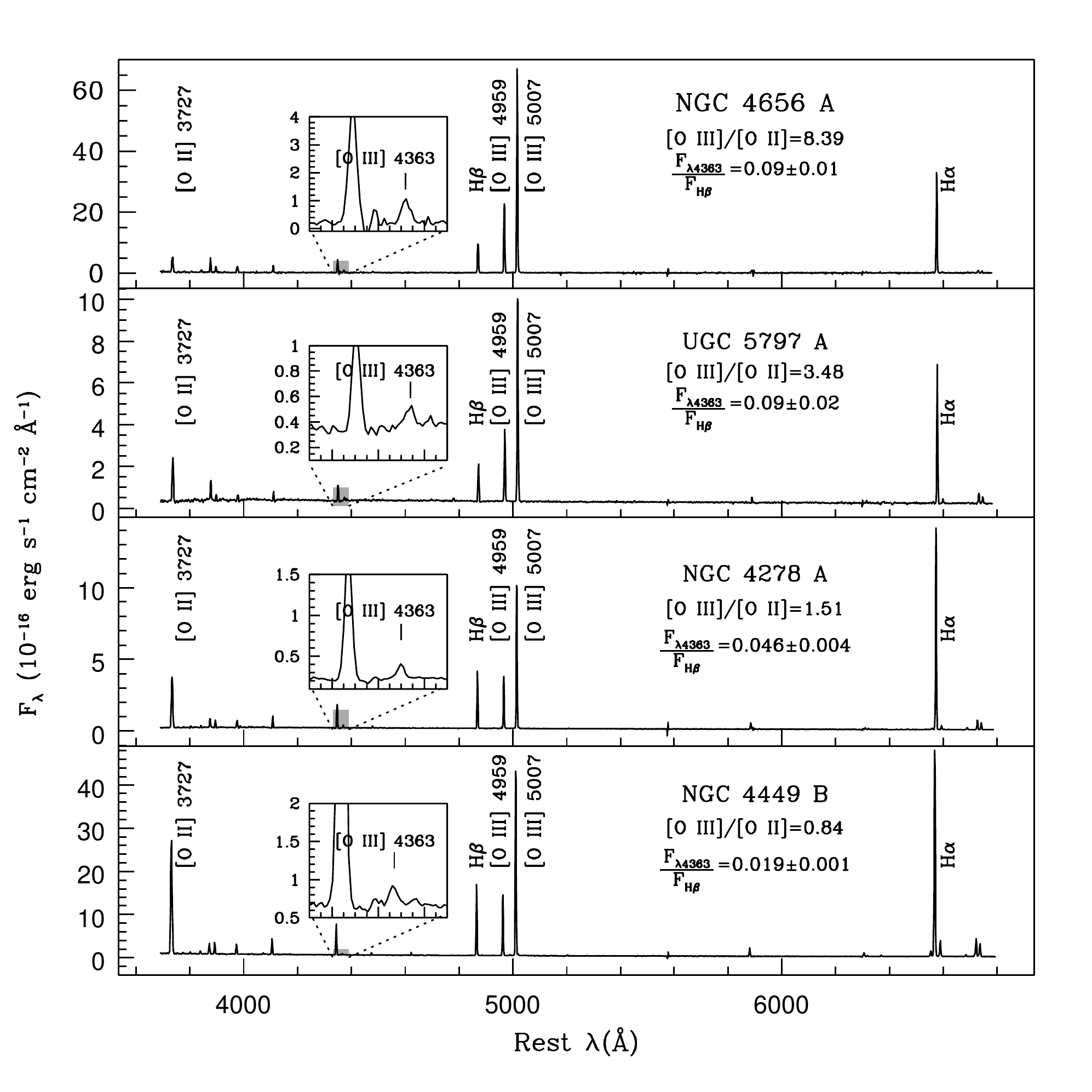}
\caption{Four sample spectra representative of the low-luminosity LVL sample presented in this paper.
The full spectral range of these high-quality, high signal-to-noise observations is shown.
The inset windows expand the region around the intrinsically faint [\ion{O}{3}] $\lambda$4363 line used to determine T$_{\rm e}$.
Note that the much stronger line blueward of [\ion{O}{3}] $\lambda$4363 is H$\gamma$ $\lambda$4340.
These spectra demonstrate the range in ionization field strength seen for this sample, ranging from
low ionization in the bottom panel (NGC 5477) to high ionization in the top panel (UGC 4656).}
\label{fig2}
\end{figure}

%----------------------------------------------------------------------------------------------------------------------

\begin{figure}
\begin{center}
\plotone{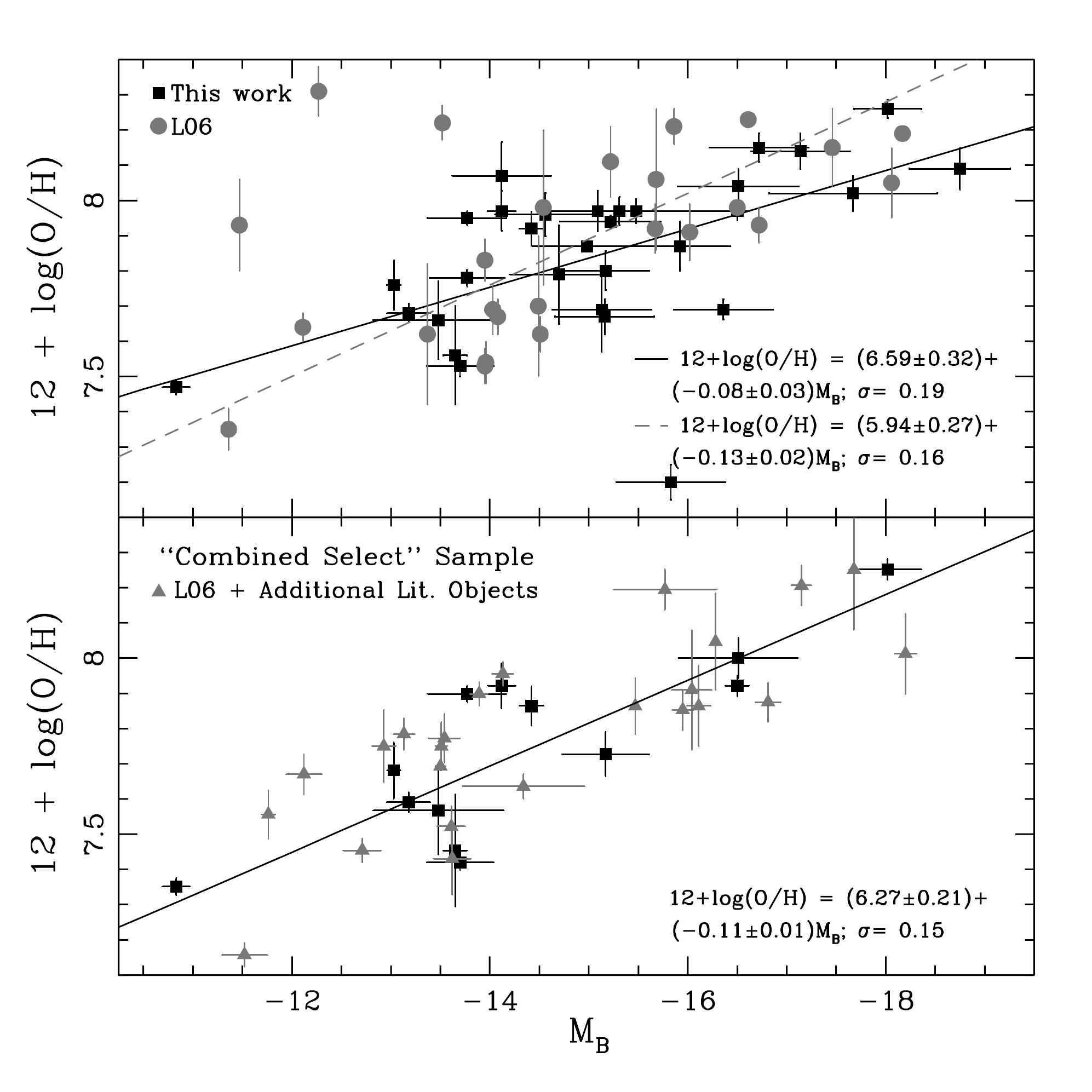}
  \caption{ On the top, the optical luminosity-metallicity relationship is plotted for the 31 objects in the present sample with ``direct" oxygen abundance measurements (squares). 
  The solid black line represents the least-squares fit to this data. 
  In comparison the original L06 dashed least-squares fit lies close to our line; in fact the slopes agree within the uncertainties. 
  The updated L06 data are plotted (which are slightly offset from the original fit - see \S 6.3 for discussion of L06 data).
  On the bottom, the optical luminosity-metallicity relationship is improved by restricting our data to a ``Combined Select" sample with ``direct oxygen" abundances and reliable distance estimates (TRGB or ceph).
  The triangles represent the set of additional ``Select" objects comprised from L06, vZ06, and \citet{marble10}, and the solid line is the least squares fit to the total ``Combined Select" sample.}
  \label{fig:LbZ}
\end{center}
\end{figure}

%----------------------------------------------------------------------------------------------------------------------

\begin{figure}
\begin{center}
\plotone{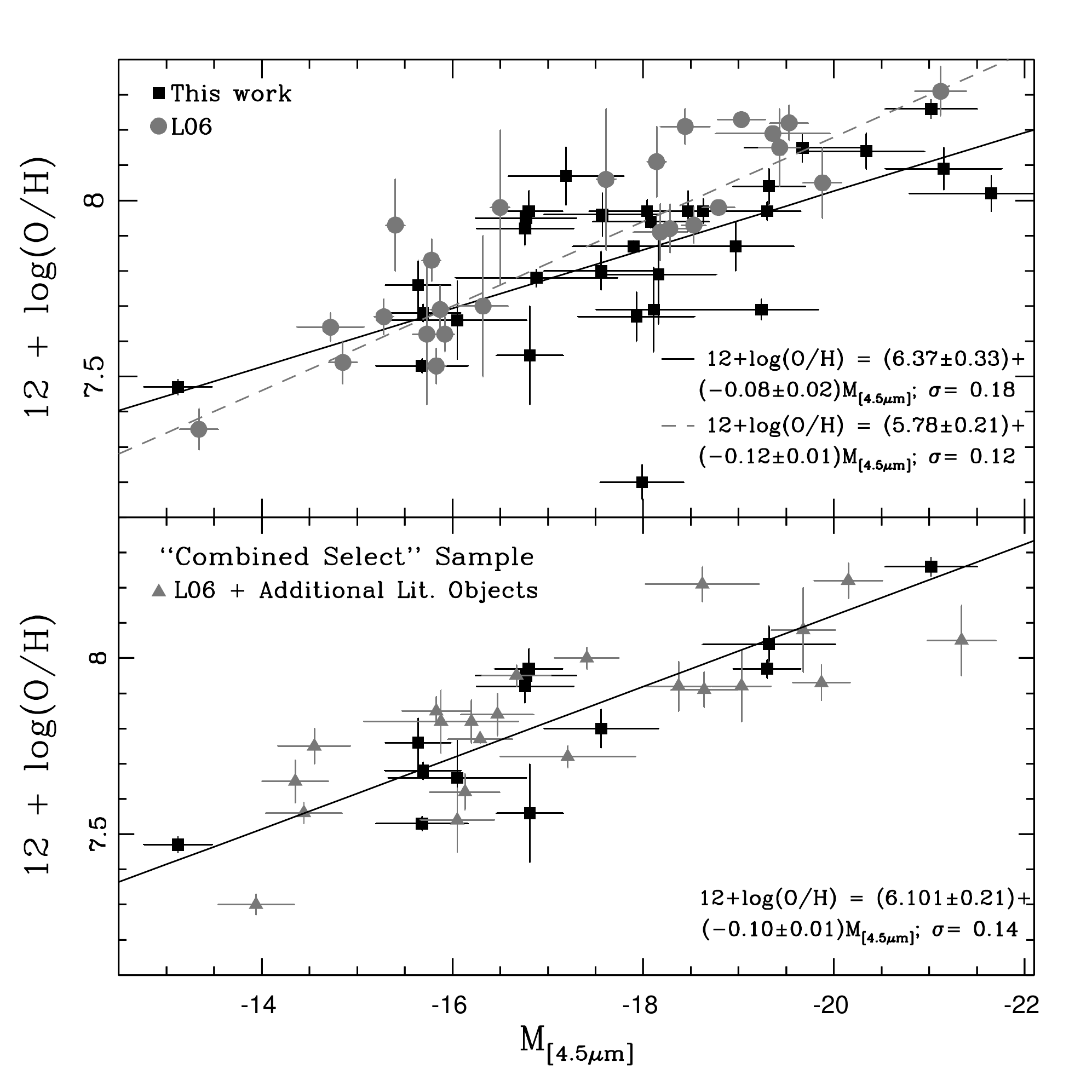}
  \caption{ In the top panel, the NIR luminosity-metallicity relationship is plotted for the 31 objects in the present sample with ``direct" oxygen abundance measurements (squares). 
  The solid black line represents the least-squares fit to this data.
  In comparison, updated L06 data are plotted (circles), with the dashed original least-squares fit of L06 lying just above our line. 
  Note that the updated data are slightly offset from the original fit (see \S 6.3 for discussion of L06 data).
  In the bottom panel is the NIR luminosity-metallicity relationship for the ``Combined Select" sample, with ``direct" oxygen abundances and reliable distance estimates (TRGB or ceph). 
  By filtering our data in this way, the L-Z relationship is strengthened.}
  \label{fig:LZ}
\end{center}
\end{figure}

%----------------------------------------------------------------------------------------------------------------------

\begin{figure}
\begin{center}
\plotone{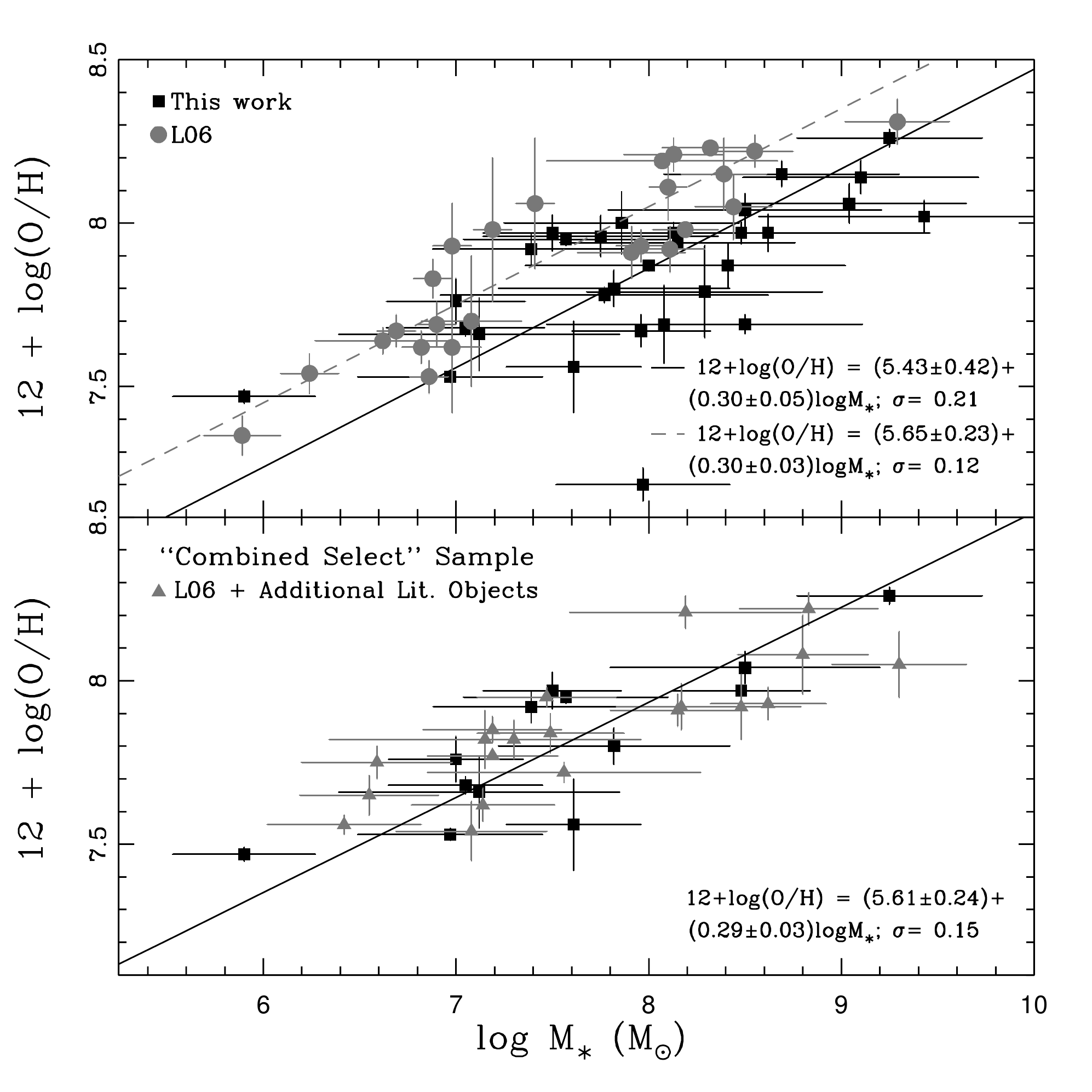}
  \caption{In the top panel the mass-metallicity relationship derived from NIR luminosities is plotted for the 31 objects in the present sample with ``direct" oxygen abundance measurements (squares). 
  The solid black line represents the least-squares fit to this data. In comparison, updated L06 data is plotted (circles). 
  We have also plotted the original least-squares fit to the sample of objects from L06 (which is not significantly offset from the updated data; see \S 6.3 for discussion of L06 data).
  This dashed gray line is offset from our estimate of the best fit. 
  In the lower panel is the mass-metallicity relationship for the ``Combined Select" sample, with ``direct" oxygen abundances and reliable distance estimates (TRGB or ceph).}
  \label{fig:MZ}
\end{center}
\end{figure}

%----------------------------------------------------------------------------------------------------------------------

\begin{figure}
  \epsscale{0.8}
  \plotone{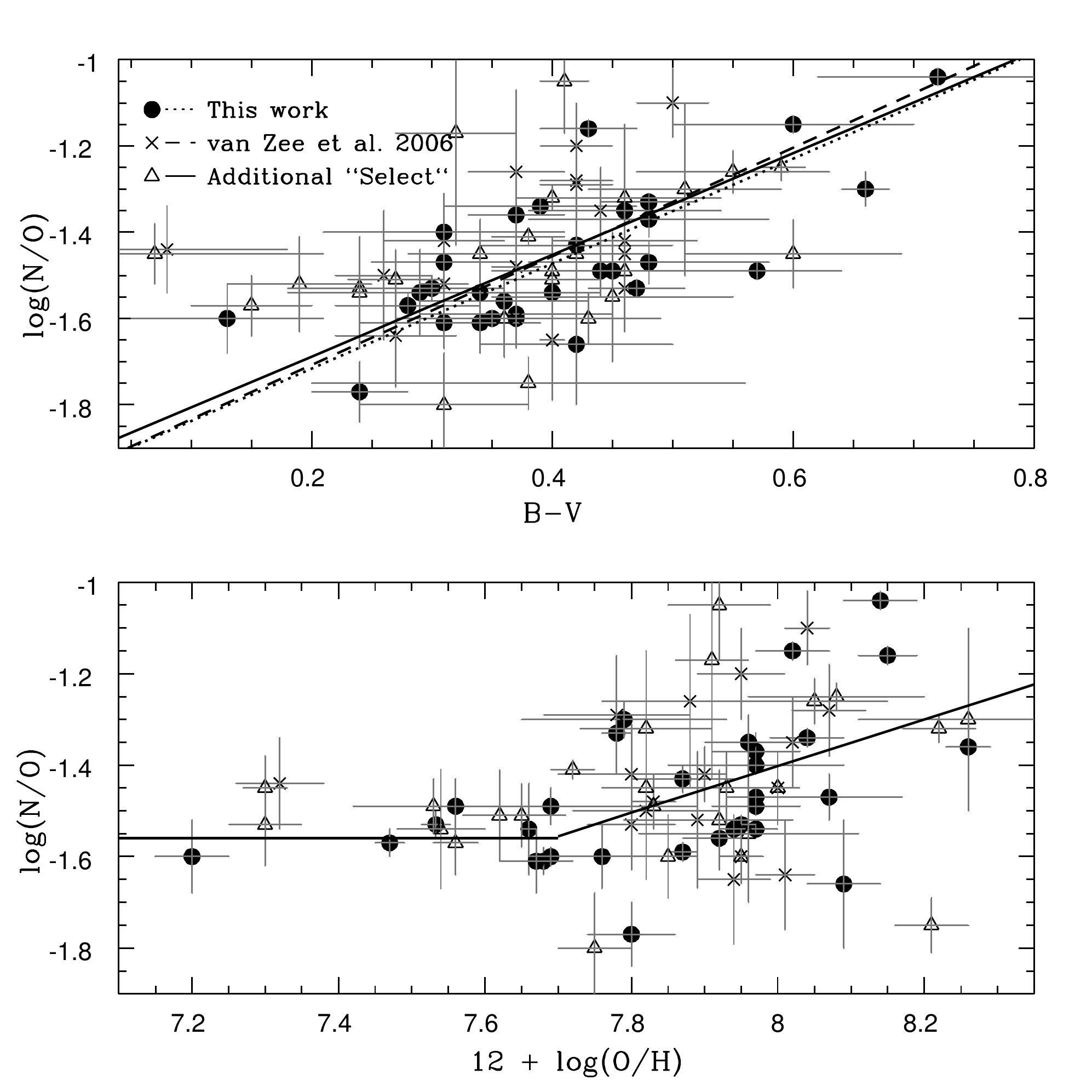}
  \caption{Relative N/O abundance is depicted. 
  The top panel compares log(N/O) to $B-V$ color for objects of the present sample with [\ion{N}{2}] observations 
  (filled circles), the sample of vZ06 (crosses), and for the additional ``Select" galaxies (triangles).
  The least squares fit for this work is represented by the dotted line.
  The dashed line is the least squares fit from vZ06.
  A solid line is shown for the literature combination of all three data sets: the present work, vZ06, and the additional values from the literature; 
  our best estimate of the true relationship for the color range of 0.05 $\lesssim B-V \lesssim$ 0.75.
  Below $B-V$ = 0.20 the data diverges from the fit, suggesting this fit is most appropriate for the range of 0.20 $\lesssim B-V \lesssim$ 0.75.
  The bottom panel displays log(N/O) versus log(O/H) for objects from this work, vZ06, and additional literature values.
  At values of 12 + log(O/H) $\le$ 7.7, the N/O values are relatively constant, consistent with the plateau associated with primary nitrogen return.
  Above this value of O/H, the scatter increases and the trend is to larger values of N/O with the addition of secondary nitrogen.}
\label{fig:NO}
\end{figure}

%----------------------------------------------------------------------------------------------------------------------

\begin{figure}
\epsscale{0.8}
\plotone{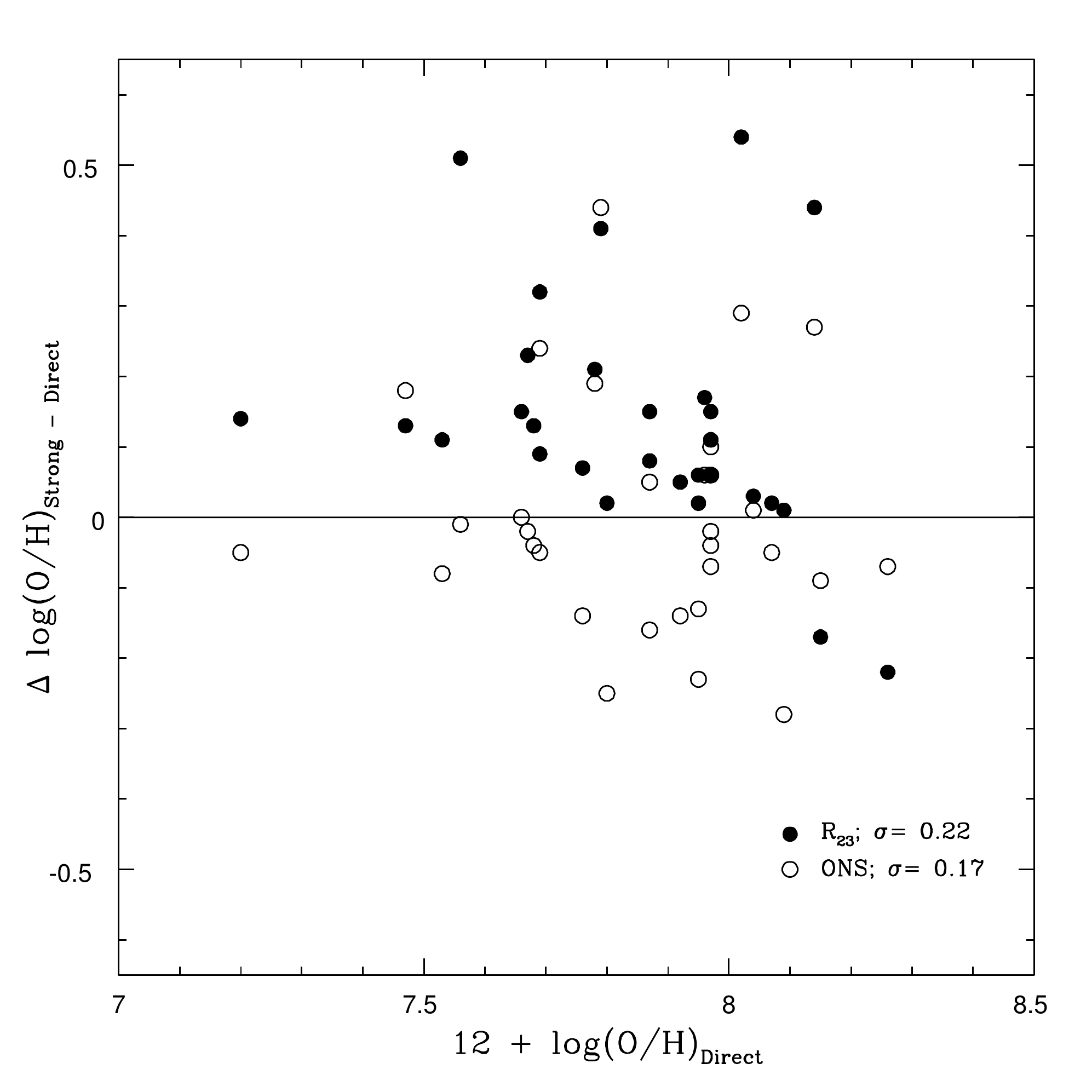}
\caption{We plot the difference in oxygen abundances determined via the ``direct" method and strong-line methods versus the 
``direct" method oxygen abundances for the objects listed in Table~\ref{tbl5}. 
The open circles display the comparison for strong-line abundances determined via the R$_{23}$ method of \citet{mcgaugh91} 
and the closed circles the ONS calibration of \citep{pilyugin10}. 
The absence of clear trends imply that simple calibrations between methods are not possible.}
\label{fig:compare}
\end{figure}

%----------------------------------------------------------------------------------------------------------------------

\begin{figure}
\epsscale{0.8}
\plotone{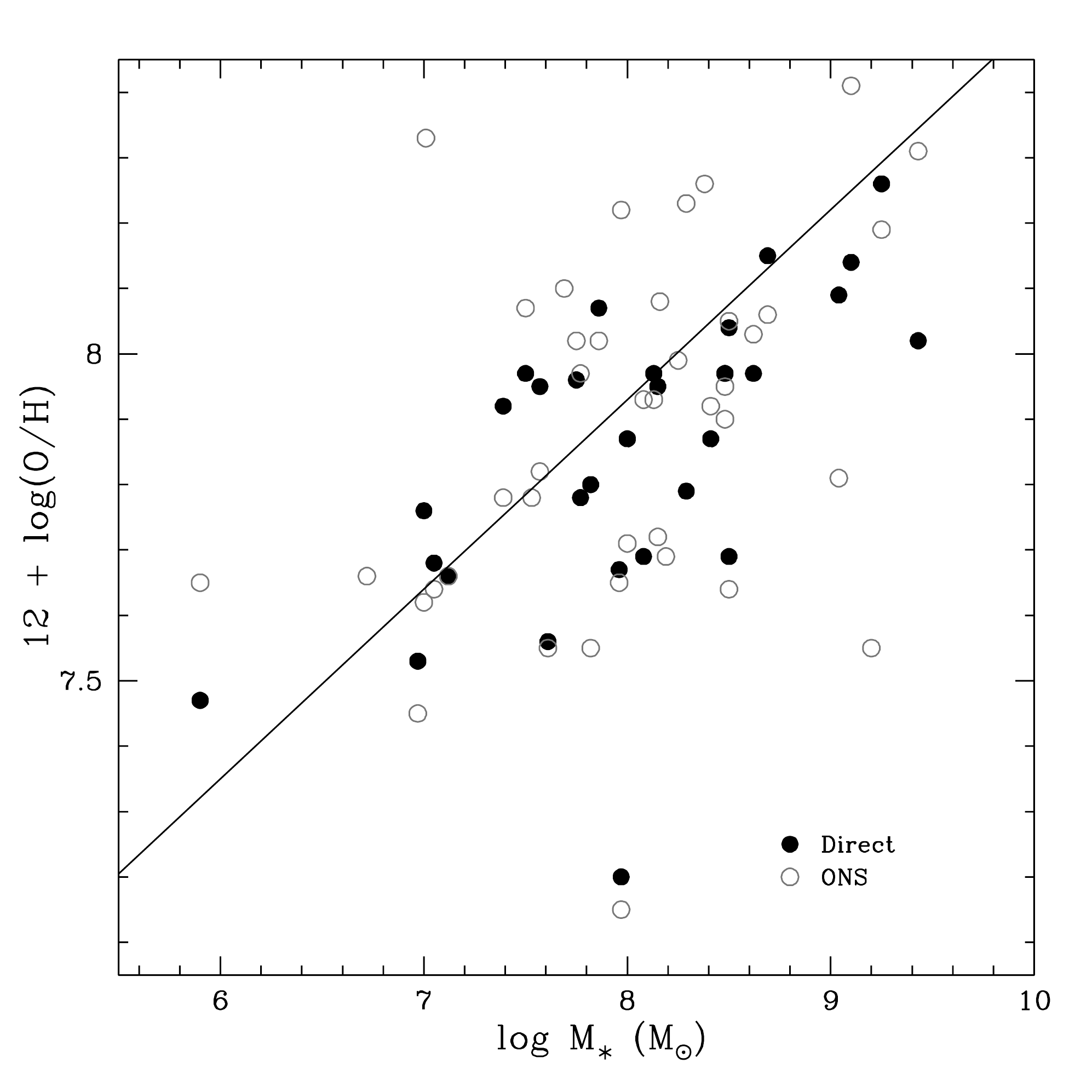}
\caption{The strong-line ONS mass-metallicity relationship is depicted for all the objects for which we calculated strong-line abundances (see Appendix~\ref{sec:strong}).
The objects are plotted in comparison to ``direct" abundances determined for the ``Combined Select" sample.
As a reference, we have plotted Equation~\ref{eqMZ}, our M-Z relationship for the ``Combined Select" sample, as a solid line.
While scatter is apparent, the overall trend is well defined and the two sample coincide.
Note that in terms of scatter the``direct" method is an improvement over the ONS strong-line calibration.}
\label{fig:strong}
\end{figure}

%----------------------------------------------------------------------------------------------------------------------

\begin{figure}
\epsscale{0.8}
\plotone{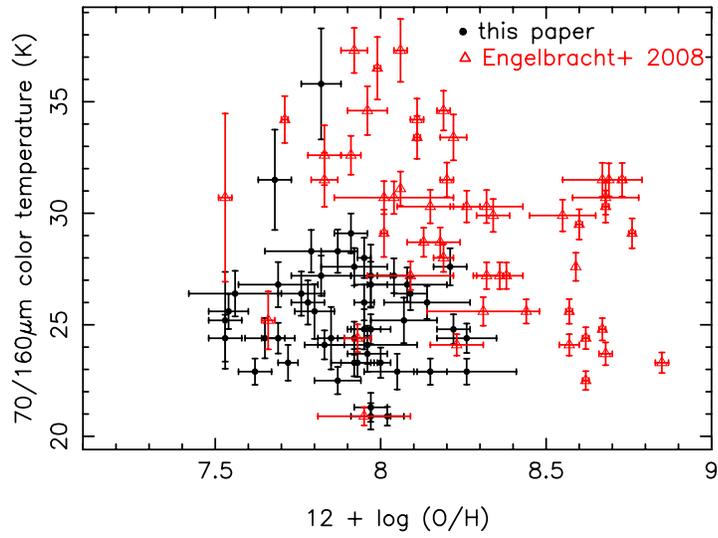}
\caption{70/160 $\mu$m color temperature versus 12 + log(O/H) is plotted for the objects presented in this paper with ``direct" oxygen abundances. 
In comparison, star-bursting galaxies from \citet{engelbracht08} with strong-line abundances seem to have larger temperatures at a given metallicity.}
\label{fig:temp}
\end{figure}

%----------------------------------------------------------------------------------------------------------------------
%----------------------------------------------------------------------------------------------------------------------

\end{document}